\documentclass[aps,pra,superscriptaddress,twocolumn,showpacs,floatfix]{revtex4-1}

\usepackage{amsmath}
\usepackage{amssymb}
\usepackage{bm}
\usepackage{graphicx}
\usepackage{appendix}
\usepackage{longtable}
\usepackage{float}
\usepackage{arydshln}
\usepackage{color}
\usepackage{color}

\begin{document}
\title{Quantum metasurfaces with periodic arrays of $\Lambda$-emitters}
%


\author{Igor V. Ryzhov}
\affiliation{Herzen State Pedagogical University, St. Petersburg, 191186 Russia}

\author{Ramil F. Malikov}
\affiliation{M. Akmullah Bashkir State Pedagogical University, 450008 Ufa, Russia}%

\author{Andrey V. Malyshev}
\affiliation{GISC, Departamento de F\'{\i}sica de Materiales,
Universidad Complutense, E-28040 Madrid, Spain}
\affiliation{Ioffe Physical-Technical Institute, 26 Politechnicheskaya str., 194021 St.-Petersburg, Russia}

\author{Victor\ A.\ Malyshev}
\affiliation{Herzen State Pedagogical University, St. Petersburg, 191186 Russia}
\affiliation{Zernike Institute for Advanced Materials, University of Groningen, Nijenborgh 4, 9747
AG Groningen, The Netherlands}

\date{\today}

\begin{abstract}
We study theoretically the optical response of a monolayer comprizing regularly spaced quantum emitters with a doublet in the ground state (the so-called $\Lambda$-emitters). The emitters' self-action through the retarded dipole-dipole interaction provides a positive feedback, interplay of which with the intrinsic nonlinearity of an isolated emitter, leads to an exotic optical dynamics of the system and prominent effects, such as multistability, self-oscillations, and quasi-chaotic behavior.
In a certain spectral domain, the monolayer operates as a bistable mirror. The optical response of the monolayer manifests high sensitivity to the doublet splitting and relaxation within the doublet, suggesting the latter to be the key parameters to tailor the monolayer optical response. These properties make such a system very promising for nanophotonic applications. We discuss the relevance of the predicted nonlinear effects for nano-sized all-optical devices.
%
%

\end{abstract}

\maketitle

\section{Introduction}
\label{Intro}
After the discovery of graphene,~\cite{Novoselov2004,CastroNeto2009} other crystalline (quasi)two-dimensional (2D) materials with fascinating optical and transport properties have been synthesized, such as transition metal dichalcogenides~\cite{ManzeliNatMatRev2017,ChernozatonskiiPhysicsUsp2018} hexagonal boron nitride, black phosphorous, and other inorganic quasi-2D systems (see for reviews Refs.~\cite{BonaccorsoMaterialsToday2012,BhimanapatiACSNano2015,TanChemRev2017}), as well as mixed-dimensional van der Waals heterostructures~\cite{JariwalaNatMat2017}.
During the last decade, artificial 2D supercrystals based on semiconductor quantum dots (SQDs)~\cite{Evers2013,Baranov2015,Ushakova2016}, organic 2D polymers~\cite{Liu2017}, and 2D nanostructures assembled from sequence-defined molecules (DNAs, peptides etc.)~\cite{MuNanoStrucNanoObjects2018} have been fabricated.

Assemblies of SQDs organized periodically are of special interest from the viewpoint of optical and opto-electronic applications, because they absorb light over the whole optical spectrum, from the infrared to the ultraviolet. These superstructures provide more degrees of freedom for manipulating their properties as compared to an isolated SQD, such as
SQD constituent, lattice geometry, and inter-dot interaction. This opens up an unprecedented possibilities to engineer physical and,
in particular, optical properties of these systems, which is offering a promising perspective for nanophotonics.

The band structure~\cite{BaimuratovSciRep2013,BaimuratovOptLett2013,BaimuratovOptLett2017,VovkPhysChemChemPhys2018} and linear optical properties~\cite{NosseMicroelectronJourn2008,BaimuratovSciRep2016} of SQD supercrystals have been addressed recently, demonstrating that the lattice geometry alone provides considerable room for engineering of the physical properties of these superstructures. Nonlinear optical response of these systems is intriguing and challenging problem and has not been widely discussed so far.

In recent publications,~\cite{MalikovEPJWebConf2017,MalyshevJPhysConfSer2019,RyzhovPRA2019} the nonlinear optical response of a 2D supercrystal comprizing SQDs with a ladder arrangement of the energy levels has been theoretically investigated. It has been found that this system can manifest fascinating nonlinear optical effects, including multistability, periodic and aperiodic self-oscillations, chaos, and transient chaos.
2D arrays of V-type quantum emitters reveal similar features~\cite{VlasovJApplSpectrosc2013,VlasovLasPhysLett2013,BayramdurdiyevEPJWebConf2019,BayramdurdiyevJETP2020}. The key ingredient to these nonlinear effects is the secondary field produced by the system dipoles, which provides an intrinsic (mirrorless) positive feedback. The interplay of the latter and the nonlinearity of the emitters themselves gives rise to the above mentioned exotic optical response of the SQD supercrystal.

\begin{figure}
\begin{center}
\includegraphics[width=0.6\columnwidth]{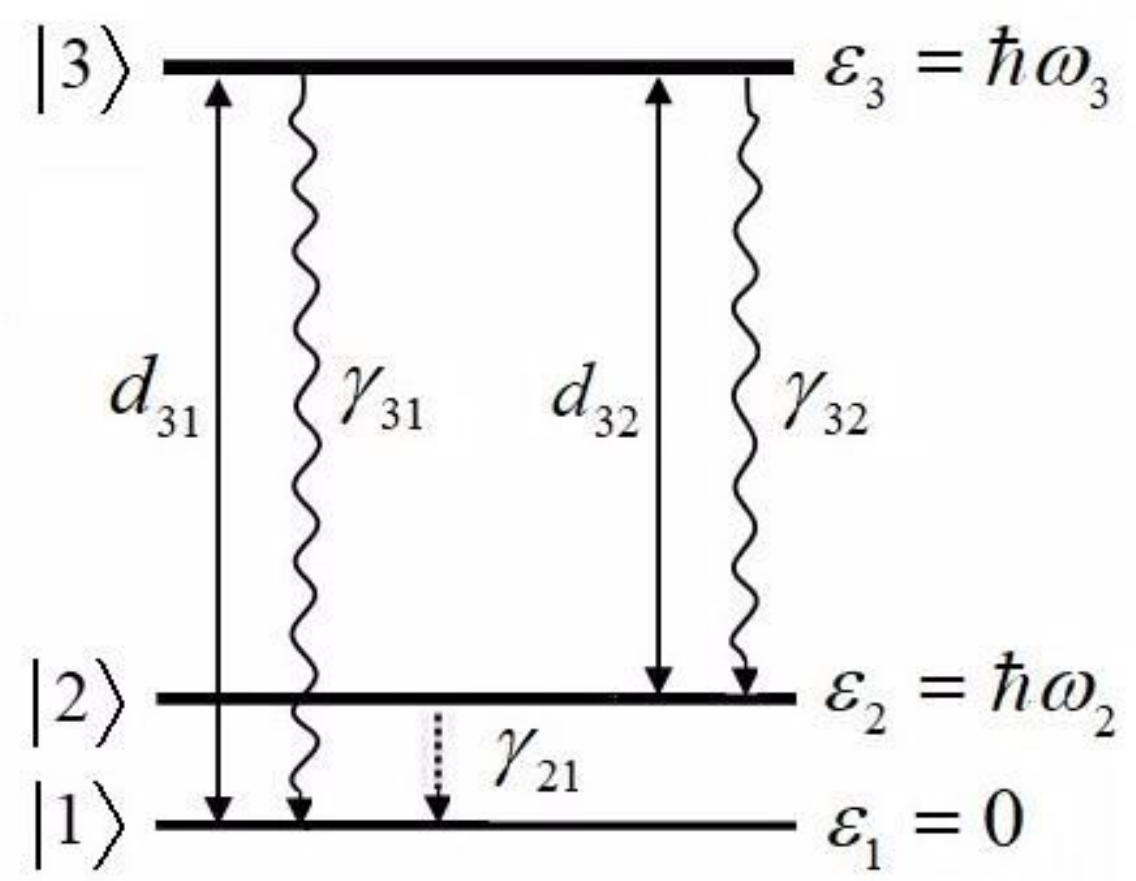}
\caption{\label{fig:Schematics} Energy level diagram of a $\Lambda$-emitter comprising the upper state $|3 \rangle$ and the lower doublet states $|2 \rangle$ and $|1 \rangle$. The energies of these states are $\varepsilon_3 = \hbar\omega_3$, $\varepsilon_2 = \hbar\omega_2$, $\varepsilon_1 = \hbar\omega_1 = 0$. Solid double-directed arrows indicate the optically allowed transitions with corresponding transition dipole moments $\bf{d}_{32}$ and $\bf{d}_{31}$. Wavy arrows denote the spontaneous decay of the upper state to the states of the doublet with rates $\gamma_{32}$ and $\gamma_{31}$. Dashed arrow indicates the relaxation within the doublet with rate $\gamma_{21}$.}
\end{center}
\end{figure}

In this paper, we investigate theoretically the nonlinear optical response of a monolayer of $\Lambda$ -emitters, i.e. emitters with a single upper state and a doublet in the ground state (see Fig.~\ref{fig:Schematics} for the level schematics). Doped quantum dots~\cite{Brunner2009} and organic nanocrystals with vibronic structure of the ground state~\cite{BookNanocrystal2011} are examples of such type of emitters. Due to a high density of emitters and high oscillator strengths of the optical transitions, the total (retarded) dipole-dipole interactions between emitters have to be taken into account, which is done in the mean-field approximation. The real part of  this interaction results in the dynamic shift of the emitter's energy levels, whereas the imaginary part describes the collective radiative decay of emitters, both depending on the population differences among levels (see, e.g., Refs.~\cite{Benedict1990,Benedict1991}). These two effects are crucial for the nonlinear dynamics of the system. As a result, in addition to bistability, analogous to that manifested by a thin layer of two-level emitters~\cite{Bowden1986,Basharov1988,Benedict1990,Benedict1991,Oraevsky1994,Malyshev2000,Glaeske2000,Klugkist2007, Malikov2017} we predict multistability, self-oscillations and quasichaotic behavior in the optical response of supercrystal of $\Lambda$-emitters. Within a certain spectral range, the monolayer operates as a bistable mirror, similar to a 2D supercrystal of SQDs with the ladder and V arrangement of the energy levels~\cite{MalikovEPJWebConf2017,MalyshevJPhysConfSer2019,RyzhovPRA2019,BayramdurdiyevEPJWebConf2019}.
To uncover the character of the instabilities, we use the standard methods of nonlinear dynamics, such as the analysis of the Lyapunov's exponents, bifurcation diagrams, phase-space maps, and Fourier spectra~\cite{AndronovBook1966,EckmannRevModPhys1985,GuckenheimerBook1986,NeimarkLandaBook1992,OttBook1993,Arnol'dBook1994,AlligoodBook1996,KatokBook1997,KuznetsovBook2004}.
To the best of our knowledge, a detailed study of the optical response of assemblies of $\Lambda$-emitters has not been carried out so far. Some preliminary results have been reported in our short paper~\cite{RyzhovEPJWebConf2019}.

The paper is organized as follows. In the next section, we describe the model of a monolayer comprizing $\Lambda$-emitters and the mathematical formalism to explore its optical response, which is based on the density matrix approach for the description of the optical dynamics of an isolated emitter and an equation for the field acting on an emitter addressed within the mean-field approximation. In Sec.~\ref{Numerics}, we present the results of numerical calculations of the monolayer optical response, including the steady-state solution (Sec.~\ref{Steady-state}), an analysis of bifurcations occurring in the system (Sec.~\ref{Bifurcation diagram}), and the system dynamics (Sec.~\ref{Dynamics}). Discussion of the underlying physical mechanisms of the predicted effects is provided in Sec.~\ref{Discussion}. In Sec.~\ref{Reflectance}, the close-to-unity monolayer reflectance is discussed. We conclude and argue about the relevance of our results for practical applications in Sec.~\ref{Summary}.

\section{Model and theoretical background}
\label{Model}
We address a 2D $N\times N$ square lattice of quantum emitters spaced by the lattice constant $a$. All emitters comprising the monolayer have the $\Lambda$-type arrangement of the energy levels (as shown in Fig.~\ref{fig:Schematics}) in which transitions are allowed only between the upper state $|3\rangle$ and a doublet $|1\rangle$ and $|2\rangle$ in the lower state. These transitions and characterized by the transition dipole moments $\bf{d}_{31}$ and $\bf{d}_{32}$, respectively. For the sake of simplicity, we set them to be real and parallel to each other, so that $\bf{d}_{32} = \mu \bf{d}_{31} \equiv \mu \bf{d}$. The upper state $|3\rangle$ decays spontaneously
to the states of the doublet $|2\rangle$ and $|1\rangle$
with the rates $\gamma_{32}$ and $\gamma_{31}$, respectively, which obey the relationship $\gamma_{32} = \mu^2\gamma_{31}$. The doublet splitting $\omega_{21}$ is assumed to be small compared to the optical transition frequencies $\omega_{32} = \omega_3 - \omega_2$ and $\omega_{31} = \omega_3 - \omega_1$. We consider also the normal incidence of the external field polarized along the transition dipole moments and having the frequency $\omega_0$.

The optical dynamics of the monolayer is governed by the Lindblad quantum master equation for the density operator $\rho(t)$~\cite{Lindblad1976,BlumBook2012}. Within the mean-field and the rotating wave approximations this equation reads
%
\begin{subequations}
\label{MasterEqAndHamiltonian}
\begin{eqnarray}
\dot{\rho}(t) = -\frac{i}{\hbar} \left[{H^\mathrm{RWA}}(t),\rho(t)\right ] + {\cal L}\{\rho(t)\}~,
\label{LindbladEq}
\end{eqnarray}
\begin{eqnarray}
H^\mathrm{RWA}(t) &=&  \hbar \left( \Delta_{21}\sigma_{22} + \Delta_{31}\sigma_{33} \right)
\nonumber\\
        &-& i\hbar\left[ \Omega_{31}(t) \sigma_{31} + \Omega_{32}(t) \sigma_{32}\right] + h.c.~,
\label{HamiltonianRWA}
\end{eqnarray}
\begin{eqnarray}
{\cal L}\{\rho(t)\} &=&
           \frac{1}{2}\gamma_{31} \left( \left[ \sigma_{13} \rho(t),
           \sigma_{31}\right]  + \left[ \sigma_{13},
           \rho(t)\, \sigma_{31}\right]\right)
           \nonumber\\
           &+& \frac{1}{2}\gamma_{32} \left( \left[
           \sigma_{23}\rho(t),\sigma_{32}\right]
           + \left[ \sigma_{23},\rho(t) \, \sigma_{32}\right]\right) \nonumber
           \nonumber\\
           &+& \frac{1}{2}\gamma_{21} \left( \left[ \sigma_{12} \rho(t),
           \sigma_{21}\right]  + \left[ \sigma_{12},
           \rho(t)\, \sigma_{21}\right]\right)~,
\label{LindbladOperator}
\end{eqnarray}
\begin{equation}
    \sigma_{ij} = |i\rangle \langle j|~, \quad i,j = 1,2,3~,
\end{equation}
\end{subequations}
%
where the dot in Eq.~(\ref{LindbladEq}) denotes the time derivative, $\hbar$ is the reduced Plank constant, $H^\mathrm{RWA}$ is the $\Lambda$-emitter Hamiltonian in the rotating wave approximation (RWA), square brackets denote commutators, ${\cal L}$ is the Lindblad relaxation operator given by   Eq.~(\ref{LindbladOperator}).~\cite{Lindblad1976,BlumBook2012}. In Eq.~(\ref{HamiltonianRWA}), $\hbar\Delta_{21} = \hbar(\omega_2 - \omega_1)$ is the doublet energy splitting, while $\hbar\Delta_{31} = \hbar(\omega_3 - \omega_0)$ is the energy of the state $|3 \rangle$ in the rotating frame. Alternatively, the quantity $\Delta_{31}$ can be interpreted as the detuning of the incident field frequency $\omega_0$ from the resonance frequency of the $1 \leftrightarrow 3$ transition. Herewith, $\Delta_{32} = \Delta_{31} - \Delta_{21}$ is the detuning of the incident field from the transition $2 \leftrightarrow 3$. Furthermore, $\Omega_{31}(t) = d_{31}E(t)/\hbar \equiv \Omega(t)$ and $\Omega_{32}(t) = d_{32}E(t)/\hbar \equiv \mu\Omega(t)$ where $E(t)$ is the slowly-varying amplitude of the total mean field acting on a $\Lambda$-emitter. The latter field is the sum of the amplitude of the incident field, $E_0(t)$, and the amplitude of the secondary field produced by all others dipoles at the position of the given $\Lambda$-emitter. Thus, $\Omega(t)$ is the Rabi amplitude of the mean field and can be written in the following form~\cite{RyzhovPRA2019} (hereafter, we omit the explicit time dependence in notations of all variables)
\begin{equation}
\label{Local field}
\Omega = \Omega_0 + (\gamma_R - i\Delta_L)(\rho_{31} + \mu\rho_{32})~,
\end{equation}
where $\Omega_0 = d_{31}E_0/\hbar$ is the Rabi amplitude of the incident field and the second term represents the Rabi amplitude of the secondary field, in which the two terms proportional to $\gamma_R$ and $\Delta_L$ are far-zone and near-zone field, respectively. The latter is analogous to the Lorentz local-field correction~\cite{BornAndWolf}. In the general case, $\gamma_R$ and $\Delta_L$ depend on the lattice geometry and the relationship between the system linear size and the excitation wavelength $\lambda$. For a simple square lattice of emitters with the linear size $Na$ one obtains~\cite{RyzhovPRA2019}
\begin{subequations}
\label{gammaRDeltaLpoint-like}
\begin{eqnarray}
\label{gammaRpoint-like}
\gamma_R = \frac{3}{8}\gamma_{31} N^2~,
\end{eqnarray}
\begin{eqnarray}
\label{DeltaLpoint-like}
\Delta_L = 3.39 \gamma_{31} \left(\frac{\lambdabar}{a}\right)^3~,
\end{eqnarray}
\end{subequations}
if $\lambda \ll Na$ (point-like system). In the opposite case of $\lambda \gg Na$ (extended system), $\gamma_R$ and $\Delta_L$ are given by~\cite{RyzhovPRA2019}
\begin{subequations}
\begin{eqnarray}
\label{gammaRextended}
\gamma_R = 4.51 \gamma_{31} \left(\frac{\lambdabar}{a}\right)^2~,
\end{eqnarray}
\begin{eqnarray}
\label{DeltaLextended}
\Delta_L  = 3.35 \gamma_{31} \left(\frac{\lambdabar}{a}\right)^3~,
\end{eqnarray}
\end{subequations}
where $\lambdabar = \lambda/(2\pi)$. As follows from Eqs.~(\ref{gammaRpoint-like}) and~(\ref{gammaRextended}), $\gamma_R$ is determined by the total number of emitters in the system for the point-like system ($\lambda \ll Na$). On the other hand, in the case of an extended sample ($\lambda \gg Na$), $\gamma_R$ is proportional to the number of emitters within the area of $\lambdabar^2$. We note here that $\gamma_R$ is nothing but the Dicke superradiant constant~\cite{Dicke1954,MalikovJETP1979,BenedictBook1996,RyzhovPRA2019} describing the collective radiation relaxation of $\Lambda$-emitters in the monolayer.

The parameter $\Delta_L$ is almost independent of the system size; it describes the near-zone dipole-dipole interaction of a given $\Lambda$-emitter with all other dipoles. Note that irrespectively of the system size, $\Delta_L \gg \gamma_R$ for a dense sample ($\lambdabar \gg a$).

In the basis of the states $|1 \rangle$, $|2 \rangle$, and $|3 \rangle$, the system of equations Eqs.~(\ref{LindbladEq}) -~(\ref{LindbladOperator}) for the density matrix elements $\rho_{\alpha\beta}$ ($\alpha,\beta = 1,2,3$) of a $\Lambda$-emitter in the monolayer reads
%
\begin{subequations}
\label{allrho}
\begin{equation}
\label{rho11}
    \dot{\rho}_{11} = \gamma_{21} \rho_{22} + \gamma_{31} \rho_{33} + \Omega^* \rho_{31} + \Omega \rho_{31}^*~,
\end{equation}
\begin{equation}
\label{rho22}
    \dot{\rho}_{22} = -\gamma_{21} \rho_{22} + \gamma_{32} \rho_{33}  + \mu(\Omega^* \rho_{32} + \Omega\rho_{32}^*)~,
\end{equation}
\begin{eqnarray}
\label{rho33}
    \dot{\rho}_{33} = &-& (\gamma_{31} + \gamma_{32}) \rho_{33} - \Omega^* \rho_{31} - \Omega \rho_{31}^*
    \nonumber\\
    &-& \mu (\Omega^* \rho_{32} + \Omega \rho_{32}^*)~,
\end{eqnarray}
\begin{eqnarray}
\label{rho31}
    \dot{\rho}_{31} = &-& \left[ i\Delta_{31} + \frac{1}{2}(\gamma_{31} + \gamma_{32}) \right] \rho_{31}
    \nonumber\\
    &+& \Omega(\rho_{33} - \rho_{11}) - \mu \Omega \rho_{21}~,
\end{eqnarray}
\begin{eqnarray}
\label{rho32}
    \dot{\rho}_{32} = &-& \left[ i\Delta_{32} + \frac{1}{2}(\gamma_{31} + \gamma_{32} + \gamma_{21}) \right] \rho_{32}
    \nonumber\\
    &+& \mu\Omega(\rho_{33} - \rho_{22}) - \Omega \rho_{21}^*~,
\end{eqnarray}
\begin{equation}
\label{rho21}
    \dot{\rho}_{21} = - \left( i\Delta_{21} + \frac{1}{2}\gamma_{21}  \right) \rho_{21} + \mu \Omega^* \rho_{31} + \Omega \rho_{32}^*~.
\end{equation}
\end{subequations}
%
Equations~\ref{rho11} -- \ref{rho21} conserve the total population, $\rho_{11} + \rho_{22} + \rho_{33} = 1$, i.e. we consider the spontaneous decay to be the only channel of the population relaxation. Dephasing of the $\Lambda$-emitter states is also neglected.

\section{Numerical results}
\label{Numerics}
\begin{figure*}[ht!]
\begin{center}
\includegraphics[width=0.7\textwidth]{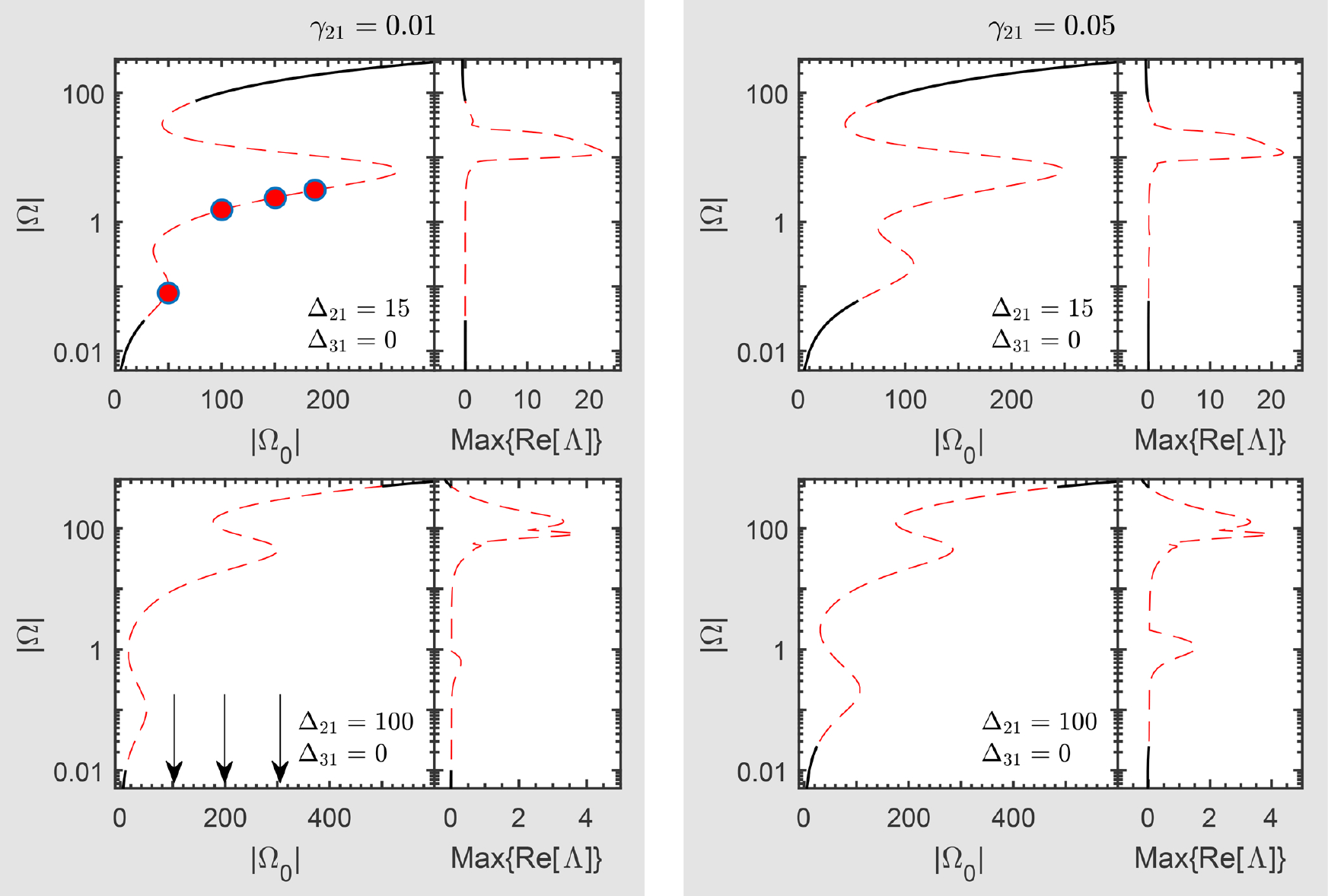}
\end{center}
\caption{\label{fig:one-photon_steady-state}
Steady-state solutions to Eqs.~(\ref{Local field}) and~(\ref{rho11})--(\ref{rho21}) obtained for the case when the incident field is in resonance with the transition $1 \leftrightarrow 3$ of an isolated $\Lambda$-emitter ($\Delta_{31} = 0$) for two values of the relaxation rate $\gamma_{21} = 0.01$ (left plot) and $\gamma_{21} = 0.05$ (right plot) and two values of the doublet splitting $\Delta_{21 } = 15$ (upper panels) and $\Delta_{21 } = 100$ (lower panels). The ratio $\mu = (\gamma_{32}/\gamma_{31})^{1/2} = 1$. Other parameters are specified in the text. Left panels in each plot display the $|\Omega|$-vs-$|\Omega_0|$ dependence. The solid (dashed) fragments of the curves indicate stable (unstable) parts of the steady-state solutions.
Circles in the upper panel (left) indicate those points of the steady-state solution, for which the time-domain behavior of $|\Omega|$ is calculated (see Fig.~\ref{fig:Delta21 = 15}). Arrows on the lower panel (left) show the Rabi magnitudes $|\Omega_0|$ of the incident field, for which the time-domain behavior of $|\Omega|$ is calculated for the initial condition $\rho_{11}(0) = 1$ (see Fig.~\ref{fig:Delta21 = 100}).
The character of a solution (stable, unstable) was unraveled from the analysis of Lyapunov's exponents $\Lambda_k$ ($k = 1,2,...8$). The maximum value of the real part of $\Lambda_k$, $\mathrm{Max}\{\mathrm{Re}[\Lambda]\}$, as a function of $|\Omega|$, are depicted in the right panels. All quantities are given in units of the radiation rate $\gamma_{31}$. }
\end{figure*}
In our numerical calculations we used the following set of parameters (similar to those in Ref.~\cite{RyzhovPRA2019}): $\gamma_{31} \approx 3\cdot 10^9$~s$^{-1}$, the ratio $\mu = d_{32}/d_{31}  = (\gamma_{32}/\gamma_{31})^{1/2}$ is taken to be unity for simplicity. The magnitudes of $\gamma_R$ and $\Delta_L$ depend on the ratio $\lambdabar/a$. Taking $\lambdabar \sim 100 \div 200$ nm and $a \sim 10 \div 20$ nm, we obtain the following estimates: $\gamma_R \sim 10^{12}$ s$^{-1}$  and $\Delta_L \sim 10^{13}$ s$^{-1}$. Accordingly, we set $\gamma_R = 100\gamma_{31}$ and $\Delta_L = 1000\gamma_{31}$. In what follows, the spontaneous emission rate $\gamma_{31}$ is used as the unit of all frequency-dimensional quantities, while $\gamma_{31}^{-1}$ is the time unit.

The remaining two parameters are the doublet splitting $\Delta_{21}$ and the relaxation rate $\gamma_{21}$ within the channel $2 \rightarrow 1$. As we show below, both affect the monolayer optical response to a large extent and, therefore, they are considered to be variable quantities. Note, in particular, that in the case of the ladder-like arrangement of energy levels (see for example, Ref.~\cite{RyzhovPRA2019}), there are no analogs of the low-frequency coherence $\rho_{21}$ and its relaxation described by the constant $\gamma_{21}$. Here, this relaxation channel provides an additional degree of freedom to tailor the optical response of the monolayer of $\Lambda$-emitters.

First, we address the monolayer optical response for the case when the incident field is in resonance with the transition $1 \leftrightarrow 3$ of an {\it isolated} $\Lambda$-emitter: $\Delta_{31} = 0, \Delta_{32} = - \Delta_{21}$. Here, it is worth to notice that the bare resonances $\Delta_{31} = 0$ and $\Delta_{32} = 0$ undergo renormalization (dressing) resulting from the action of the secondary field on a given $\Lambda$-emitter. This field not only shifts the energy levels, but also couples transitions $1 \leftrightarrow 3$ and $2 \leftrightarrow 3$ to each other. As the result, one of the dressed resonances remains approximately around $\Delta_{31} = 0$, while the other one shifts down to $\Delta_{31} \approx \Delta_L$ (see Sec.~\ref{Discussion} for more detailed discussion). The monolayer optical response in the vicinity of $\Delta_{31} = \Delta_L$ will be considered as well (Sec.~\ref{Reflectance}).

The system of equations (\ref{Local field}) and~(\ref{rho11})--(\ref{rho21}) is a system of stiff equations, characterized by several significantly different time scales. In our case, they are $\gamma_{31}^{-1}\gg \gamma_R^{-1} \gg \Delta_L^{-1}$. The doublet splitting $\Delta_{21}$ and the relaxation rate $\gamma_{21}$ bring in two time scales more. To solve the system numerically we use specialised routines for stiff equations.
%

\subsection{Steady-state analysis}
\label{Steady-state}
To begin with, we consider the steady-state regime, setting to zero all time derivatives in Eqs.~(\ref{rho11})--(\ref{rho21}). We use the analytical method developed in our recent paper~\cite{RyzhovPRA2019} to solve these steady-state equations. The results, calculated for two values of the doublet splitting $\Delta_{21}$ and two relaxation rates $\gamma_{21}$, are presented in Fig.~\ref{fig:one-photon_steady-state}.

As is seen from Fig.~\ref{fig:one-photon_steady-state}, the Rabi magnitude $|\Omega|$ of the mean field can have several solutions (up to five for $\gamma_{21} = 0.05$ and $\Delta_{21} = 15$) for a given value of the Rabi magnitude $|\Omega_0|$ of the incident field. To explore the stability of different solutions, we used the standard Lyapunov's exponents analysis~\cite{NeimarkLandaBook1992,OttBook1993}. calculating the eigenvalues $\Lambda_k$ ($k=1\ldots 8$) of the Jacobian matrix of the right hand side of Eqs.~(\ref{rho11})--(\ref{rho21}) as a function of $|\Omega|$. The Lyapunov's exponent $\Lambda_k$ with the maximal real part, $\mathrm{Max}_k\{\mathrm{Re}[\Lambda_k]\}$, determines the character of a given steady-state solution: if $\mathrm{Max}_k\{\mathrm{Re}[\Lambda_k]\} \leq 0$ the solution is stable and unstable otherwise. The values of $\mathrm{Max}_k\{\mathrm{Re}[\Lambda_k]\}$ are plotted in the right part of each panel in Fig.~\ref{fig:one-photon_steady-state}.

We point out a very important feature of the steady-state characteristics: not only branches with the negative slope are unstable, which is always the case, but those with the positive slopes as well. Quite remarkably also, both the lower and upper branches are only partially stable. In particular, for $\Delta_{21} = 100$, the steady-state curve is almost completely unstable within the considered range of Rabi magnitudes $|\Omega_0|$. The character of instabilities of different branches of the steady-state solutions is discussed in the next section.

\subsection{Bifurcation diagram}
\label{Bifurcation diagram}
\begin{figure}[ht!]
\begin{center}
\includegraphics[width=0.7\columnwidth]{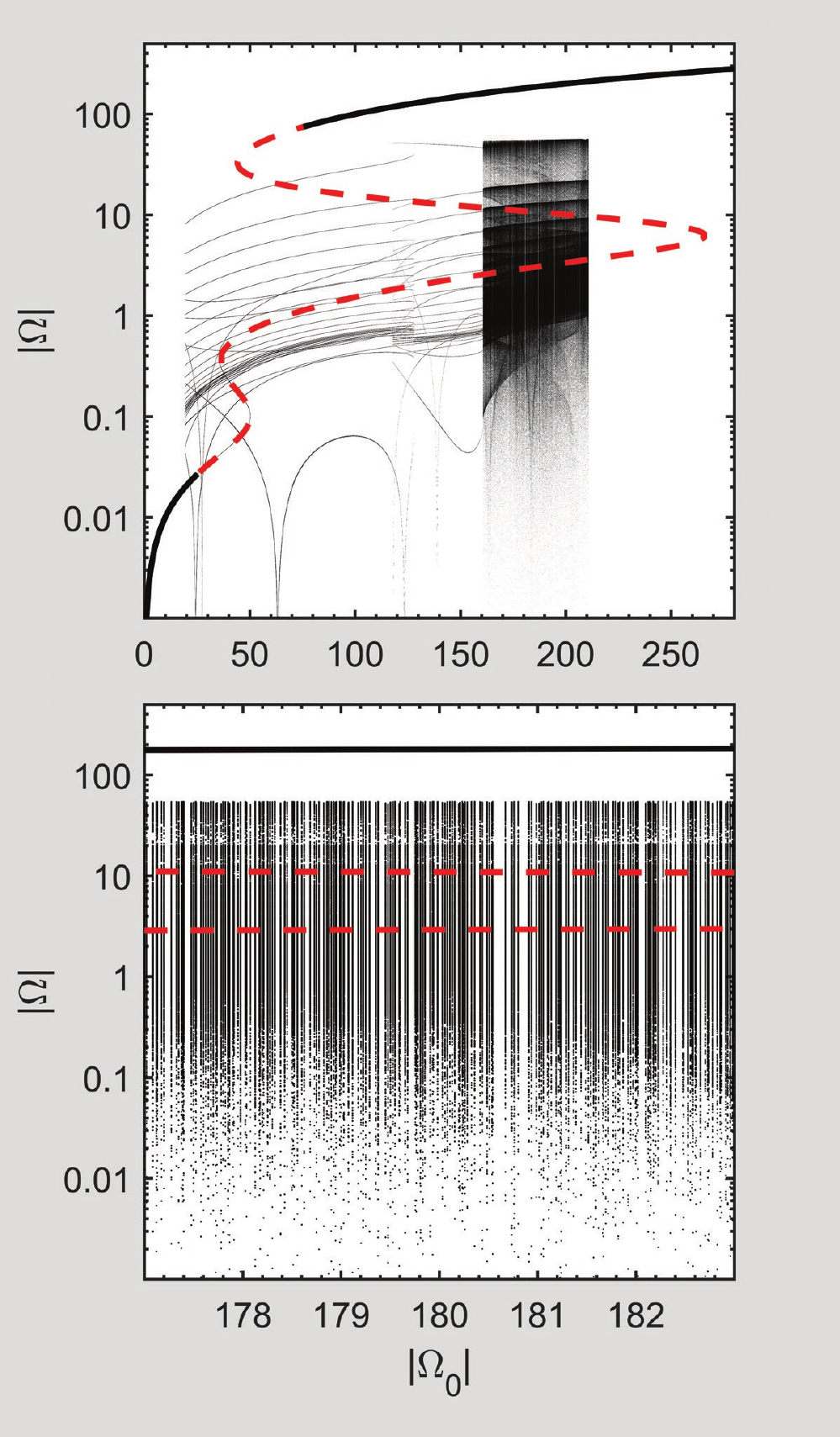}
\end{center}
\caption{\label{fig:One-photon bifurcation diagram Delta21 = 15}
Top: the overall bifurcation diagram (extrema of the Rabi magnitude $|\Omega(t)|$ of the mean field as a function of the Rabi magnitude $|\Omega_0|$) of the incident field calculated for the case when the incident field is in resonance with the transition $1 \leftrightarrow 3$ of an isolated $\Lambda$-emitters ($\Delta_{31} = 0$) setting the doublet splitting $\Delta_{21} = 15$, the relaxation rate $\gamma_{21} = 0.01$, and the ratio $\mu = (\gamma_{32}/\gamma_{31})^{1/2} = 1$. Other parameters are specified in the text. The steady-state (double S-shaped) solution is given for reference. Bottom: the blow-up of the dark regions of the bifurcation diagram with nontrivial dynamics. All quantities are given in units of the radiation rate $\gamma_{31}$. }
\end{figure}
\begin{figure}[ht!]
\begin{center}
\includegraphics[width=0.71\columnwidth]{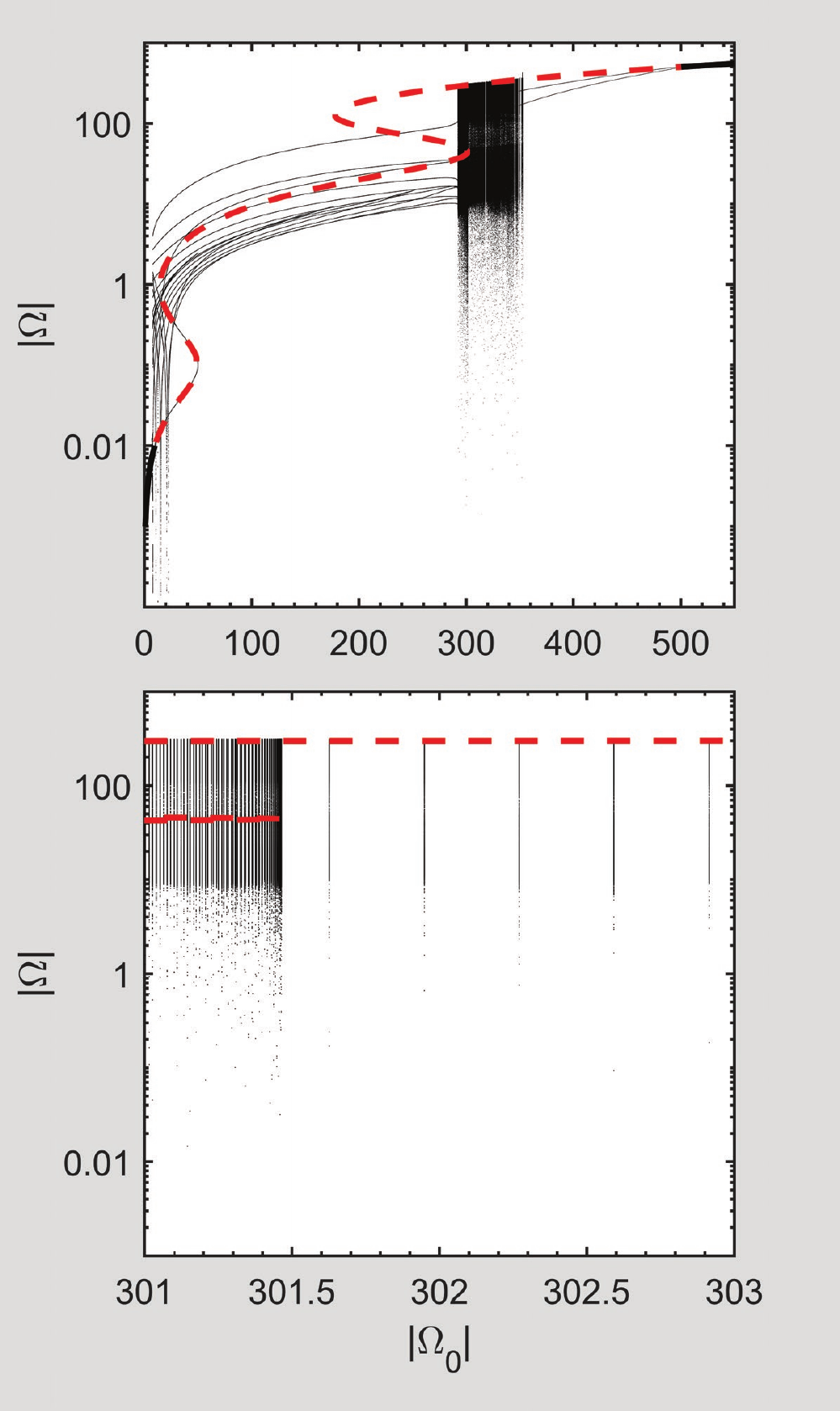}
\end{center}
\caption{\label{fig:One-photon bifurcation diagram Delta21 = 100}
Same as in Fig.~\ref{fig:One-photon bifurcation diagram Delta21 = 15}, but for $\Delta_{21} = 100$.}
\end{figure}
The bifurcation diagram is a powerful tool to explore possible scenarios of a dynamical system behavior~\cite{GuckenheimerBook1986,NeimarkLandaBook1992,Arnol'dBook1994,KuznetsovBook2004}. The diagram represents the system dynamics qualitatively as a function of some controlling (bifurcation) parameter. In our case, the most natural bifurcation parameter is the Rabi magnitude $|\Omega_0|$ of the incident field, while the Rabi magnitude of the mean field $|\Omega|$ is one of the possible measurable outputs. Below, we describe briefly an efficient procedure of constructing the bifurcation diagram, which was proposed in Ref.~\cite{RyzhovPRA2019}.
%

For each value of $|\Omega_0|$ and appropriate sets of initial conditions, Eqs.~(\ref{rho11})--(\ref{rho21}) are integrated until all transients vanish and the system reaches an attractor. Then the dynamics is analysed in the following way: all the extrema of $|\Omega(t)|$ are obtained over a sufficiently long time interval $T$. All the extremal values of $|\Omega(t)|$ are plotted as points for the current value of the Rabi magnitude $|\Omega_0|$, forming the bifurcation diagram. The distribution of the extrema provides qualitative information on possible types of the system dynamics. For example, if the dynamics converges to a stable fixed point, all the extrema collapse onto a single point given by the corresponding stable steady-state solution. If the system is on a periodic orbit, all the extrema collapse onto a finite set of points separated by gaps.  Quasiperiodic motion would appear as vertical bars separated by gaps, while chaos would manifest itself as a continuous vertical line. For more detail, see Ref.~\cite{RyzhovPRA2019}.

Figures~\ref{fig:One-photon bifurcation diagram Delta21 = 15} and~\ref{fig:One-photon bifurcation diagram Delta21 = 100} show the bifurcation diagrams, calculated as described above, for the case when the incident field is tuned into the resonance $1 \leftrightarrow 3$ ($\Delta_{31} = 0, \Delta_{32} = - \Delta_{21}$) of the isolated $\Lambda$-emitter. The data presented in Fig.~\ref{fig:One-photon bifurcation diagram Delta21 = 15}  are obtained for the doublet splitting $\Delta_{21} = 15$, while those in Fig.~\ref{fig:One-photon bifurcation diagram Delta21 = 100} -- for $\Delta_{21} = 100$. The relaxation constant $\gamma_{21} = 0.01$ in both cases. The double-S-shaped steady-state characteristics (solid black and dashed red curves), which are taken from Fig.~\ref{fig:one-photon_steady-state}, are given here for reference.
Upper panels in both figures display the overall bifurcation diagram, while the lower ones show blow ups of the dense features in the diagrams.

The trivial part of the bifurcation diagrams coincides with the stable part of the corresponding steady-state curves (stable fixed points). However, there are also parts of the bifurcation diagrams which reveal nontrivial dynamics. On the one hand, there are sets of thin lines of points separated by gaps corresponding to some periodic orbits (limit cycles). On the other hand, there are dense features formed by vertical very dense lines of points, suggesting that the extrema of the mean-field Rabi magnitude $|\Omega(t)|$ are randomly distributed, so that the signal might be of the chaotic nature. Thus, for all values of the Rabi magnitude $|\Omega_0|$ of the incident field, the system dynamics manifests various types of attractors. We address their properties in more details below (see Sec.~\ref{Dynamics}).

Figures~\ref{fig:One-photon bifurcation diagram Delta21 = 15} and~\ref{fig:One-photon bifurcation diagram Delta21 = 100} show that the system undergos multiple bifurcations. Consider, for example, the vicinity of $|\Omega_0| \approx 20$ in Fig.~\ref{fig:One-photon bifurcation diagram Delta21 = 15} and assume that the system is initially at the lower stable branch of the steady state (fixed point). As $|\Omega_0|$ increases the system would follow the lower branch until it looses stability at some value of $|\Omega_0|$ and a stable limit cycle is born. This scenario is known as a supercritical Andronov-Hopf bifurcation. If the field is swept back, the system stays at the limit cycle attractor, until it disappears, converting back to the stable fixed point, which resembles a subcritical Andronov-Hopf bifurcation~\cite{GuckenheimerBook1986,Arnol'dBook1994,KuznetsovBook2004}. Note that the two changes of system dynamics occur at different values of $|\Omega_0|$, so the bifurcation diagram demonstrates hysteresis.

The next bifurcation is about $|\Omega_0 \approx 160|$ (left side of the dense feature), where on sweeping up $|\Omega_0|$, a limit cycle converts into a chaotic trajectory, thus resembling a subcritical Andronov-Hopf bifurcation~\cite{GuckenheimerBook1986,Arnol'dBook1994,KuznetsovBook2004}. The system remains on the chaotic attractor if $|\Omega_0|$ is increased slowly enough until (about $|\Omega_0 \approx 210|$) the chaotic attractor disappears and the system collapses onto a stable fixed point (the upper stable branch of the stationary solution).

Here, it is helpful to mention that the chaos found may turn out to be of transient nature, in the sense that, if the system is let to evolve for sufficiently long time, it will finally be attracted to one of the stable steady-state points. Such evens can be seen in the blow-up shown in the lower panel of Fig.~\ref{fig:One-photon bifurcation diagram Delta21 = 15}: the white gaps in the feature correspond to solutions that converged during the calculation time towards the stable steady-state solution. However, the time required in order that transients are over is hardly predictable. Besides, this time seems to be very sensitive to initial conditions and the integration method, which is a typical feature of a transient chaos (see Refs.~\cite{Lai2011,Tel2015} and references therein).
%
%
%
\begin{figure*}[ht!]
\begin{center}
\includegraphics[width=0.8\textwidth]{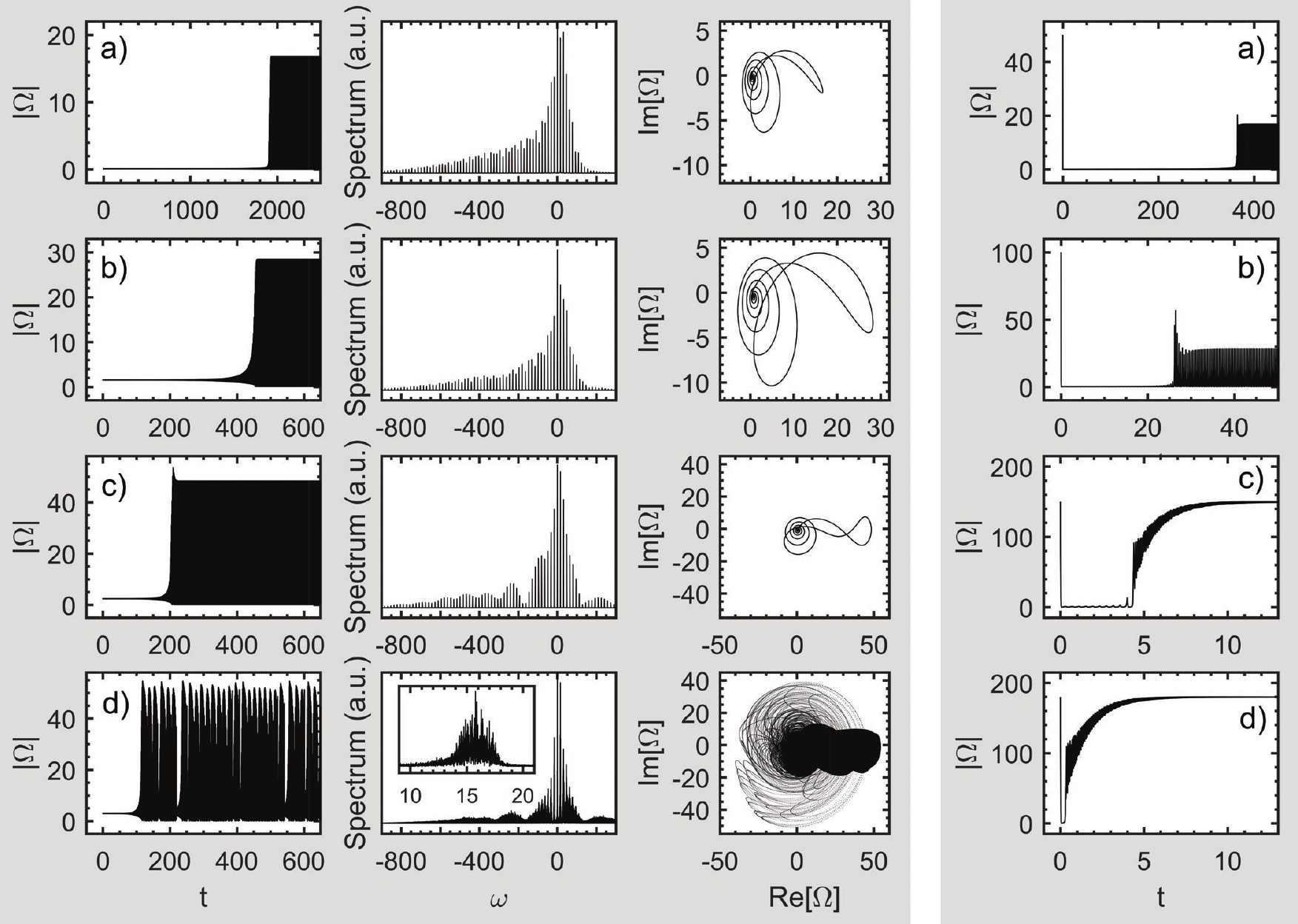}
\end{center}
\caption{\label{fig:Delta21 = 15}
Left plot - Time-domain behavior of the Rabi magnitude $|\Omega|$ of the mean field (left panels), the Fourier spectrum $|\int_{T} \exp{i\omega t} \Omega(t) dt|$ (middle panels), and the two-dimensional phase-space map ($\mathrm{Re}[\Omega], \mathrm{Im}[\Omega]$) of the attractor (right panels) obtained by solving Eqs.~(\ref{rho11})--(\ref{rho21}) for the doublet splitting $\Delta_{21} = 15$. The ratio $\mu = (\gamma_{32}/\gamma_{31})^{1/2} = 1$. Other parameters are specified in the text.
Four points on the steady-state characteristics were considered as the initial conditions: a) - ($|\Omega_0| = 50, |\Omega| = 0.0923$), b) - ($|\Omega_0| = 100, |\Omega| = 1.5213$), c) - ($|\Omega_0| = 150, |\Omega| = 2.3637$), and d) - ($|\Omega_0| = 180, |\Omega| = 2.9251$). Right plot - Time-domain behavior of the Rabi magnitude $|\Omega|$ of the mean field calculated for the ground state as the initial condition, $\rho_{11}(0) = 1$, at the same values of Rabi magnitude $|\Omega_0|$ of the incident field as in the left plot: a) - $|\Omega_0| = 50$, b) - $|\Omega_0| = 100$, c) - $|\Omega_0| = 150$, and d) - $|\Omega_0| = 180$. The inserts blow up the details of the Fourier spectrum. All frequency-dependent quantities are given in units of the radiation rate $\gamma_{31}$, while time is in units of $\gamma_{31}^{-1}$. }
\end{figure*}
\subsection{Dynamics}
\label{Dynamics}
\begin{figure*}[ht!]
\begin{center}
\includegraphics[width=0.7\textwidth]{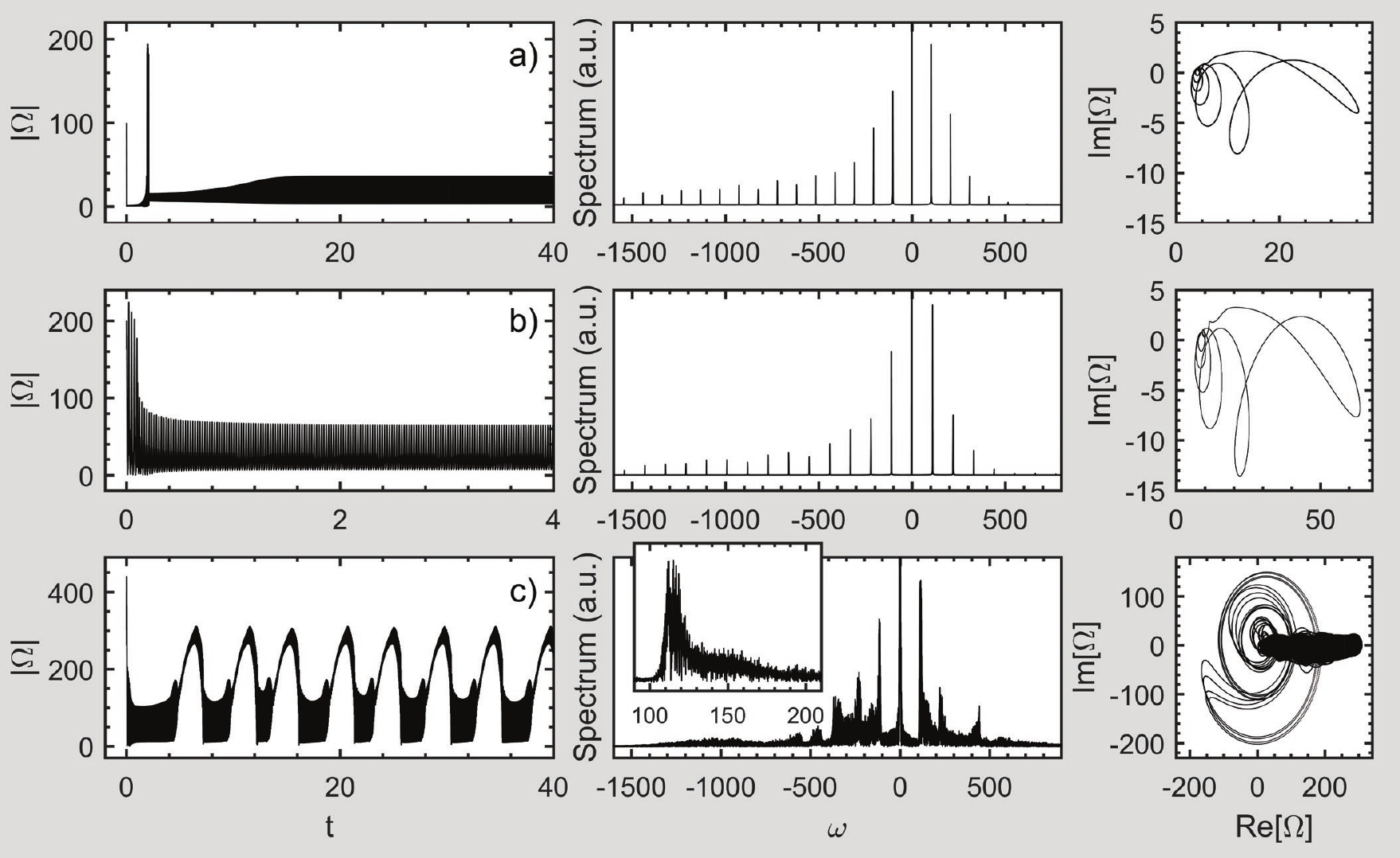}
\caption{\label{fig:Delta21 = 100}
Time-domain behavior of the Rabi magnitude $|\Omega|$ of the mean field (left panels), the Fourier spectrum $|\int_{T} \exp{i\omega t} \Omega(t) dt|$ (middle panels), and the two-dimensional phase-space map ($\mathrm{Re}[\Omega], \mathrm{Im}[\Omega]$) of the attractor (right panels) obtained by solving Eqs.~(\ref{rho11})--(\ref{rho21}) with the ground state as the initial condition, $\rho_{11}(0) = 1$, setting the doublet splitting $\Delta_{21} = 100$ and the ratio $\mu = (\gamma_{32}/\gamma_{31})^{1/2} = 1$. Other parameters are specified in the text. Three values of the Rabi magnitude $|\Omega_0|$ of the incident field were used in calculations: a) - $|\Omega_0| = 100$, b) - $|\Omega_0| = 200$, c) - $|\Omega_0| = 301.4$. The inserts blow up the details of the Fourier spectrum. All frequency-dimension quantities are given in units of the radiation rate $\gamma_{31}$, while time is in units of $\gamma_{31}^{-1}$. }
\end{center}
\end{figure*}
In this section, we present a variety of nontrivial attractors which the system achieves in its evolution. When solving Eqs.~(\ref{rho11})--(\ref{rho21}), two types of initial conditions were considered: first, when the system initially is in the ground state ($\rho_{11}(0) = 1$, while all other density matrix elements are equal to zero), and second, when the system starts from a given point of the steady-state solution. Here, it is worthwhile to note that if the system is exactly in a steady state, it remains there forever. However, due to the finite precision of the initial state itself, the system is, in fact, in a very small vicinity of the exact steady state. Therefore, it is either attracted back to the initial state (if the latter is a stable fixed point) or drifts away from it otherwise.

The results of numerical calculations are presented in Fig.~\ref{fig:Delta21 = 15} (for the doublet splitting $\Delta_{21} = 15$) and Fig.~\ref{fig:Delta21 = 100} (for $\Delta_{21} = 100$). Shown in the left plot of Fig.~\ref{fig:Delta21 = 15} are: time-domain behavior of the mean-field magnitude $|\Omega(t)|$ (left panels), the Fourier spectrum $|\int_{T} \exp{i\omega t} \Omega(t) dt|$ (middle panels) and two-dimensional phase-space map ($\mathrm{Re}[\Omega], \mathrm{Im}[\Omega]$) of the attractor (right panels) for four points on the steady-state characteristics chosen as the initial conditions (shown by thick dots in Fig.~\ref{fig:one-photon_steady-state}, upper panel): a) - ($|\Omega_0| = 50, |\Omega| = 0.0923$), b) - ($|\Omega_0| = 100, |\Omega| = 1.5213$), c) - ($|\Omega_0| = 150, |\Omega| = 2.3637$), and d) - ($|\Omega_0| = 180, |\Omega| = 2.9251$). We observe that for the points a) - c) the system evolves towards limit cycles, which has its confirmation in the equidistant character of the attractor's Fourier spectrum and in closedness of the attractor's trajectory. Oppositely, for the point d), residing within the black feature of the bifurcation diagram, the attractor resembles a chaotic motion: its Fourier spectrum is of a quasi-continuous nature and the trajectory densely covers a finite area in the phase space.

The right plot in Fig.~\ref{fig:Delta21 = 15} shows the results of calculations performed for the case when the system resides initially in the ground state, $\rho_{11}(0) = 1$, for the same set of the incident field magnitude $|\Omega_0|$ as in the left plot: a) - $|\Omega_0| = 50$, b) - $|\Omega_0| = 100$, c) - $|\Omega_0| = 150$, and d) - $|\Omega_0| = 180$. In first two cases, a) and b), the system's attractors appear to be insensitive to the initial conditions, whereas for the rest, c) and d), the system ends up on the upper stable branch of the steady-state characteristics. Sensitivity to the initial conditions is an inherent property of dynamical systems and is commonly considered as one of the possible way to encode information~\cite{Gao2008}.

Figure~\ref{fig:Delta21 = 100} shows the results of numerical calculations of the mean-field dynamics performed for the doublet splitting $\Delta_{21} = 100$ and the ground-state initial conditions: $\rho_{\alpha\beta}(0) = \delta_{\alpha 1}\delta_{\beta 1}$. The panels are ordered in the same way as in Fig.~\ref{fig:Delta21 = 15} (left plot). Three values of the Rabi magnitude $|\Omega_0|$ of the incident field were considered (shown by arrows in Fig.~\ref{fig:Delta21 = 15}, left plot, lower panel): a) - $|\Omega_0| = 100$, b) - $|\Omega_0| = 200$, c) - $|\Omega_0| = 301.4$.

As is seen from Fig.~\ref{fig:Delta21 = 100}, for $|\Omega_0| = 100$ and $|\Omega_0| = 200$, the system, during its evolution, is attracted to limit cycles, whereas for $|\Omega_0| = 301.4$, the attractor is rather a chaotic orbit. Choosing the initial conditions on the steady-state curve (for the same values of $|\Omega_0|$) changes only a little the character of attractors in this case (not shown).

\subsection{Qualitative reasoning}
\label{Discussion}
As was shown above, the monolayer optical response demonstrates a variety of fascinating properties: multistability, self-oscillations, and chaos. The origin of such a behavior is derived from the secondary field produced by the emitters place of a given one, which depends on the current state of the latter. This provides a positive feedback resulting finally in instabilities. On neglecting the secondary field, all above mentioned features disappear.

To specify our reasoning, consider Eqs.~(\ref{rho31}) and~(\ref{rho32}). Substituting  therein the expression (\ref{Local field}) for the mean-field Rabi amplitude $\Omega$, we get
%
%
\begin{subequations}
\begin{eqnarray}
\label{R31 extended}
    \dot{\rho}_{31} = &-& \left[ i\Delta_{31} + \Gamma_{31} \right.
    \nonumber\\
    &-& \left. (\gamma_R - i\Delta_L)(Z_{31} -  \mu\rho_{21})\right] \rho_{31}
    \nonumber\\
    &+& \mu(\gamma_R - i\Delta_L)(Z_{31} -\mu\rho_{21}) \rho_{32}
    \nonumber\\
    &+& \Omega_0 (Z_{31} - \mu\rho_{21})~,
\end{eqnarray}
\begin{eqnarray}
\label{R32 extended}
    \dot{\rho}_{32} = &-& \left[ i\Delta_{32} + \Gamma_{32} \right.
    \nonumber\\
    &-& \left. \mu(\gamma_R - i\Delta_L)(\mu Z_{32} - \rho_{21}^*)\right] \rho_{32}
    \nonumber\\
    &+& (\gamma_R - i\Delta_L)(\mu Z_{32} - \rho_{21}^*) \rho_{31}
    \nonumber\\
    &+& \Omega_0 (\mu Z_{32} - \rho_{21}^*)~,
\end{eqnarray}
\end{subequations}
where we denoted $\Gamma_{31} = \frac{1}{2}(\gamma_{31} + \gamma_{32})$, $\Gamma_{32} = \frac{1}{2}(\gamma_{31} + \gamma_{32} + \gamma_{21})$, $Z_{31} = \rho_{33} - \rho_{11}$, and $Z_{32} = \rho_{33} - \rho_{22}$. Equations~(\ref{R31 extended}) and~~(\ref{R32 extended}) are nothing else than those describing two coupled nonlinear oscillators driven by two incident forces. An important peculiarity of this system is that all oscillator's characteristics (frequencies, relaxation rates, coupling strengths, and driving forth amplitudes) depend on the current state of a given $\Lambda$-emitter. The latter originates direct from the secondary field acting on the $\Lambda$-emitter on the part of the others (these do not appear in analogous equations for an isolated emitter). The principal consequence of this action is twofold: it results, first, in a renormalization of the transition frequencies, that is $\omega_{31} \rightarrow \omega_{31} + \Delta_L Z_{31} + \mu {\cal I}m[(\gamma_R - i\Delta_L)\rho_{21}]$ and $\omega_{32} \rightarrow \omega_{32} + \mu^2\Delta_L Z_{21} + \mu {\cal I}m[(\gamma_R - i\Delta_L)\rho_{21}^*]$ for transitions $1 \leftrightarrow 3$ and $2 \leftrightarrow 3$, respectively, and, second, in an additional damping of the corresponding transitions described by $-\gamma_R Z_{31} + \mu {\cal R}e[(\gamma_R - i\Delta_L)\rho_{21}]$ and $-\mu^2\gamma_R Z_{32} + \mu {\cal R}e[(\gamma_R - i\Delta_L)\rho_{21}^*]$.
We stress again that both renormalizations depend on the current state of the emitter, thus having a {\it dynamic} nature.

Before the incident field is switched on, $Z_{31}(0) = -1$, whereas $Z_{32}(0) = \rho_{21}(0) = 0$, the latter is because the states $|2 \rangle$ and $|3 \rangle$ are not populated. Accordingly, only the transition $1 \leftrightarrow 3$ experiences the above mentioned renormalization, while the others do not. So the starting conditions are: the actual detuning away from the resonance with the transition $1 \leftrightarrow 3$ and the relaxation rate of the latter acquire values $\Delta_{31} - \Delta_L$ and $(1/2)(\gamma_{31} + \gamma_{32}) + \gamma_R$, respectively. As $\Delta_L \gg \Delta_{31}$ and $\gamma_R \gg (1/2)(\gamma_{31} + \gamma_{32})$, namely $\Delta_L$ and $\gamma_R$ determine the resonance detuning and the relaxation rate of the $1 \leftrightarrow 3$ transition at the initial instant. All other resonance detunings and decay rates remain unchanged.

The second terms in the right-hand sides of Eqs.~(\ref{R31 extended})  and~(\ref{R32 extended}) couple the oscillators to each other through the secondary field: $\rho_{31}$ to $\rho_{32}$ with the coupling strengths $(\gamma_R - i\Delta_L)(Z_{31} - \mu\rho_{21})$ and $\rho_{32}$ to $\rho_{31}$ with the strength $(\gamma_R - i\Delta_L)(\mu Z_{32} - \rho_{21}^*)$. Initially, the oscillators are decoupled because $Z_{32}(0) = \rho_{32}(0) = \rho_{21}(0) = 0$. However, they do couple as soon as the upper doublet state $|2\rangle$ is populated, which occurs immediately after population of the emitter's higher state $|3\rangle$ and subsequent decay to the upper state  $|2\rangle$ of the doublet. Interconnection of the transitions $2\leftrightarrow 1$ and $3\leftrightarrow 2$ results in an additional {\it dynamic} coupling-driven renormalization of the transition frequencies and relaxation rates. In what follows, we will refer to the whole secondary-field-driven renoormalization of $\Lambda$-emitter's states/transitions as to dressing of $\Lambda$-emitter.

We believe that a complicated interplay of the underlined dynamic dressing effects, changing the resonance conditions and subsequent population redistribution among levels drive finally the system to being unstable.

\section{Reflectance}
\label{Reflectance}
In our analysis of the monolayer optical response we used as an output the Rabi amplitude $\Omega$ of the mean field. In experiment, however, the reflected or transmitted field is commonly measured. These two are determined by the far-zone part of $\Omega$ and are given by the following expressions:
\begin{subequations}
\begin{equation}
\label{Reflected field}
\Omega_\mathrm{refl} = \gamma_R (\rho_{31} + \mu\rho_{32})~.
\end{equation}
\begin{equation}
\label{Transmitted field}
\Omega_\mathrm{tr} = \Omega_0 + \gamma_R (\rho_{31} + \mu\rho_{32})~.
\end{equation}
\end{subequations}
The reflectance $R$ and transmittance $T$ (reflection and transmission coefficients of the light flow, respectively) are then defined as
\begin{equation}
\label{Reflectance and Transmittance}
R = \left|\frac{\Omega_\mathrm{refl}}{\Omega_0}\right|^2, \quad T = \left|\frac{\Omega_\mathrm{tr}}{\Omega_0}\right|^2~.
\end{equation}
\begin{figure*}[ht!]
\begin{center}
\includegraphics[width=0.8\linewidth]{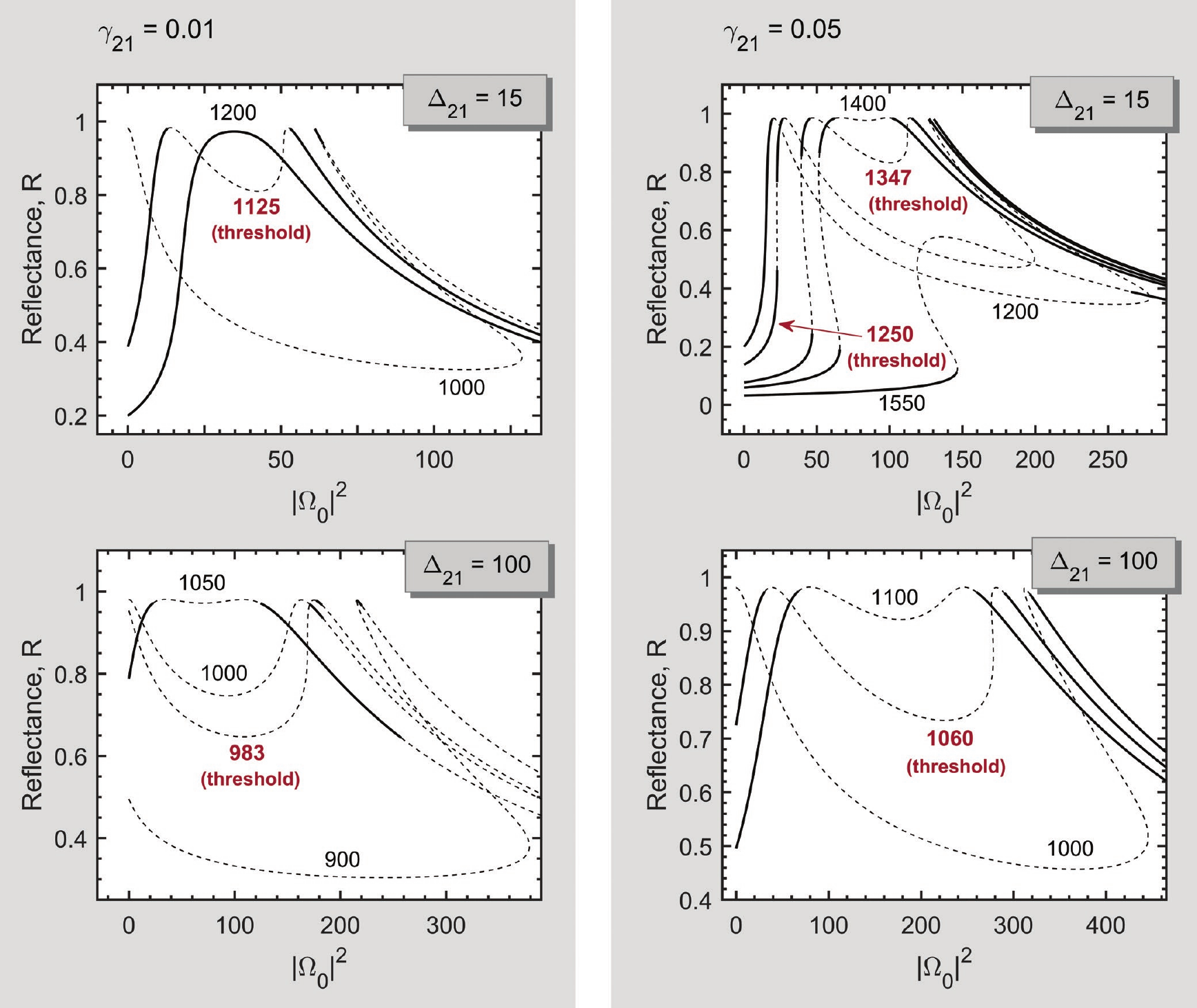}
\end{center}
\caption{\label{fig:Reflectance Delta21 = 100} Intensity dependence of the steady-state reflectance $R$ calculated for the detuning $\Delta_{31}$ in the vicinity of $\Delta_{31} = \Delta_L$ for two values of the relaxation rate $\gamma_{21} = 0.01$ (left plot) and $\gamma_{21} = 0.05$ (right plot) and two values of the doublet splitting $\Delta_{21 } = 15$ (upper panels) and $\Delta_{21 } = 100$ (lower panels). The values of $\Delta_{31}$ considered are shown in the plots. The ratio $\mu = (\gamma_{32}/\gamma_{31})^{1/2} = 1$. Other parameters are specified in the text. The solid (dotted) fragments of the curves indicate their stable (unstable) parts. All frequency-dimension quantities are given in units of the radiation rate $\gamma_{31}$, while $|\Omega_0|^2$ is in units of $\gamma_{31}^2$.  }
\end{figure*}
\begin{figure*}[ht!]
\begin{center}
\includegraphics[width=0.9\linewidth]{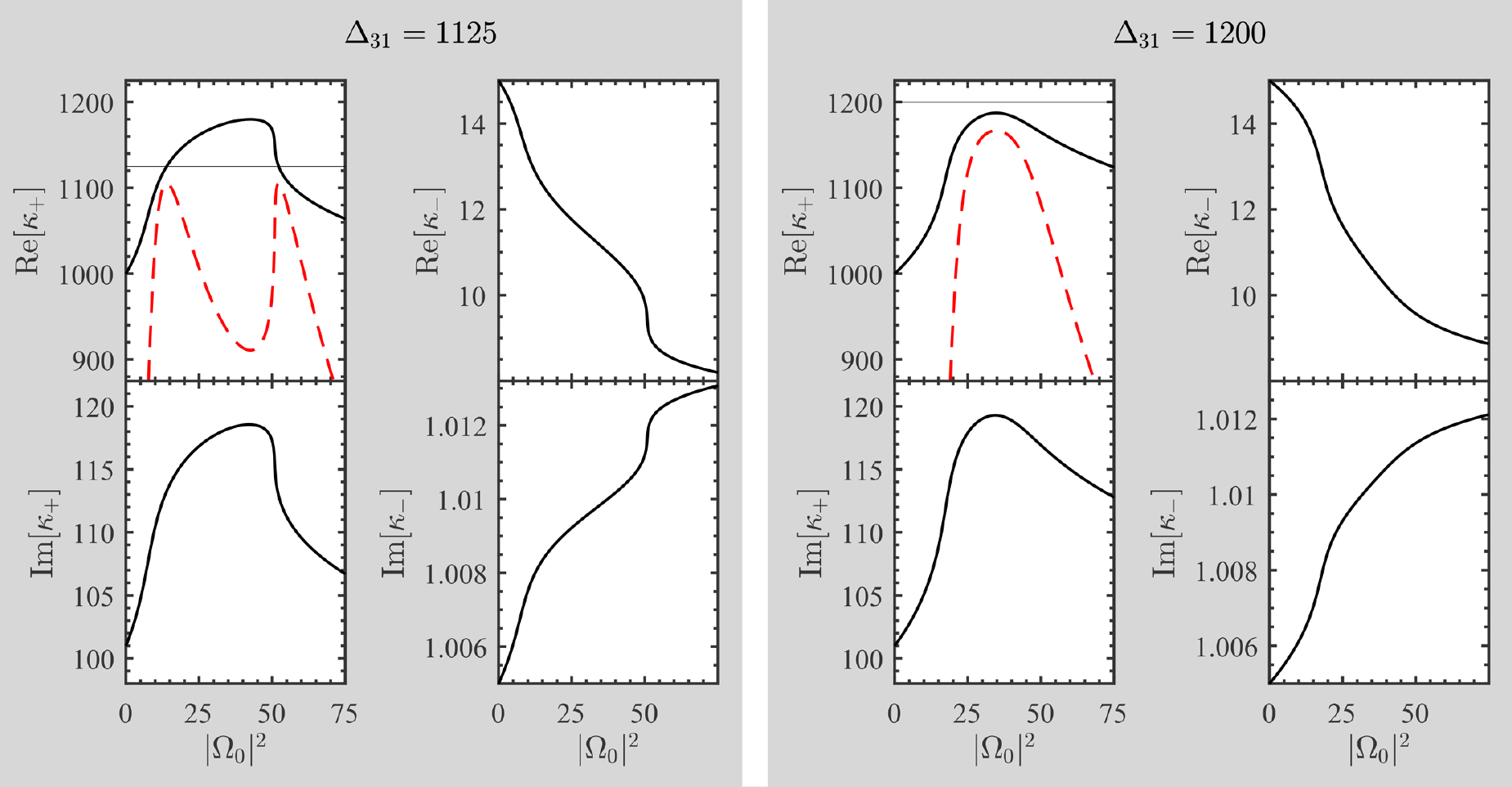}
\end{center}
\caption{\label{fig:kappa plus} Intensity dependence of zeros $\kappa_{\pm}$ of the denominator in Eq.~(\ref{rho31+mu rho32}) (solid curves) calculated for two values of the detuning $\Delta_{31}$; left plot - $\Delta_{31} = 1125$, right plot - $\Delta_{31} = 1200$. Dashed curves show the reflectance $R$ scaled to the plot (given for reference). Horizontal thin lines denote $\Delta_{31}$-level. Parameters of the doublet are: $\Delta_{21} = 15$, $\gamma_{21} = 0.01$. The ratio $\mu = (\gamma_{32}/\gamma_{31})^{1/2} = 1$. Other parameters are specified in the text. All frequency-dimension quantities are given in units of $\gamma_{31}$, while $|\Omega_0|^2$ is in units of $\gamma_{31}^2$.  }
\end{figure*}
\begin{figure*}[ht!]
\begin{center}
\includegraphics[width=0.8\textwidth]{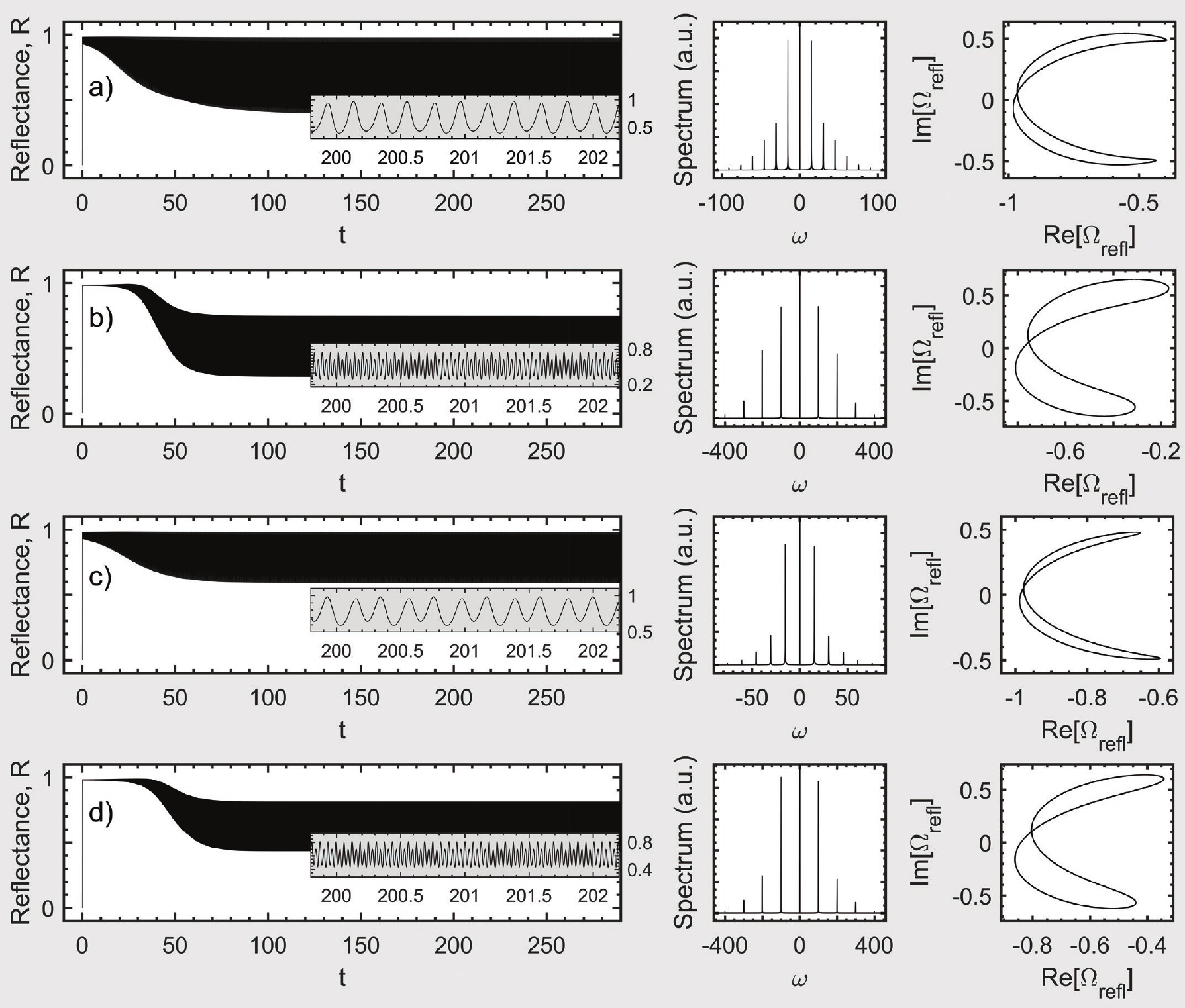}
\end{center}
\caption{\label{fig:Reflectance dynamics Delta21 = 15} Time-domain behavior of the reflectance $R$ (left panels), the Fourier spectrum $|\int_{T} \exp{i\omega t} \Omega_\mathrm{refl}(t) dt|$ (middle panels) and the two-dimensional phase-space map ($\mathrm{Re}[\Omega_\mathrm{refl}], \mathrm{Im}[\Omega_\mathrm{refl}]$) of the attractor (right panels) obtained by solving Eqs.~(\ref{rho11})--(\ref{rho21}) for the detuning $\Delta_{31} = 1000$ with the ground-state initial condition, $\rho_{11}(0) = 1$, and setting the Rabi magnitude of the incident field $|\Omega_0| = 1$. The other parameters are: a) -- $\Delta_{21} = 15$, $\gamma_{21} = 0.01$; b) -- $\Delta_{21} = 100$, $\gamma_{21} = 0.01$; c) -- $\Delta_{21} = 15$, $\gamma_{21} = 0.05$; d) -- $\Delta_{21} = 100$, $\gamma_{21} = 0.05$. The inserts blow up the details of dynamics. All frequency-dependent quantities are given in units of the radiation rate $\gamma_{31}$, while time is in units of $\gamma_{31}^{-1}$. }
\end{figure*}

\subsection{Steady-state}
\label{Reflectance steady-state}
Look first at the linear regime of excitation ($|\Omega_0|\ll 1$) and restrict ourselves to analyzing the steady-state reflectance. At $|\Omega_0| \ll 1$, the major contribution to $\Omega_\mathrm{refl}$ comes from $\rho_{31}$ which is given by
\begin {equation}
\label{eq:rho21 linear}
    \rho_{31} = - \frac{\Omega_0}{ i(\Delta_{31} - \Delta_L) + \frac{1}{2}(\gamma_{31} + \gamma_{32}) + \gamma_R}~.
\end{equation}
Substituting Eq.~(\ref{eq:rho21 linear}) into Eq.~(\ref{Reflectance and Transmittance}), for the reflectance $R$ we get
\begin{equation}
\label{Reflectance linear}
    R = \frac{\gamma_R^2}{(\Delta_{31} - \Delta_L)^2 + \left[\frac{1}{2}(\gamma_{31} + \gamma_{32}) + \gamma_R\right]^2}~.
\end{equation}
From this expression, it follows that for the range of detuning $\Delta_{31} \le 100$ used in our calculations so far, the reflectance $R \approx (\gamma_R/\Delta_L)^2 \ll 1$, because $\Delta_L \gg \Delta_{31}, \gamma_R, \gamma_{31} + \gamma_{32}$ for the parameters of our model. Nevertheless, all features of the mean field ($\Omega$) found in Sec.~\ref{Numerics} will translate into the reflected field ($\Omega_\mathrm{refl}$) and thus can be observed in the reflection geometry.

By contrast, in the vicinity of the renormalized by the near field resonance, $\Delta_{31} \sim \Delta_L$ (genuine resonance), the reflectance is close to unity, $R \approx 1$, i.e., in this region of frequencies, the monolayer of $\Lambda$-emitters operates as {\it a perfect reflector}.
%
%

Now, address the nonlinear regime of reflectance ($|\Omega_0| \gg 1$) in the vicinity of $\Delta_{31} \sim \Delta_L$, where the linear reflectance is high, Figure~\ref{fig:Reflectance Delta21 = 100} shows the results of numerical calculations of the intensity-dependent reflectance,  $R$-vs-$|\Omega_0|^2$, for two values of the relaxation rate $\gamma_{21} = 0.01$ (left plot) and $\gamma_{21} = 0.05$ (right plot) and two values of the doublet splitting $\Delta_{21 } = 15$ (upper panels) and $\Delta_{21 } = 100$ (lower panels). From this figure, one observes that, for the selected values of the detuning $\Delta_{31}$, $R$-vs-$|\Omega_0|^2$ behavior is non-monotonous, can be multivalued and unstable (dashed parts of curves). Moreover, some of solutions are unstable in the whole range of intensities considered, like, for example, those for $\Delta_{31} = 1000$. One more feature of the nonlinear reflectance is that the $R$-vs-$|\Omega_0|^2$ dependence is mostly double-peaked and, what is surprising, the peaks match up the total reflection, $R \approx 1$, i.e., on increasing the intensity of the incident field, the system falls twice in resonance with the latter.

In spite of the fact that we are now in the truly nonlinear regime, the problem in part can be handled in-depth. Solving formally Eqs.~(\ref{rho31}) and~(\ref{rho32}) together with Eq.~(\ref{Local field}) in the steady-state regime [${\dot\rho}_{31}(t) = {\dot\rho}_{31}(t) =0$], we can find $\rho_{31} + \mu\rho_{32}$ which determine the Rabi amplitude $\Omega_\mathrm{refl}$ of the reflected field, Eq.~(\ref{Reflected field}), and after that the reflectance $R$.. The result reads
\begin{widetext}
\begin{eqnarray}
\label{rho31+mu rho32}
    R = \gamma_R^2 \left| \frac{\mu(i\Delta_{31} + \Gamma_{31})(\mu Z_{32} - \rho_{21}^*)
    + (i\Delta_{32} + \Gamma_{32})(Z_{31} - \mu\rho_{21})}
    {(i\Delta_{31} + \Gamma_{31})(i\Delta_{32} + \Gamma_{32})
    -(\gamma_R - i\Delta_L)[\mu(i\Delta_{31} + \Gamma_{31})(\mu Z_{32} - \rho_{21}^*)
    + (i\Delta_{32} + \Gamma_{32})(Z_{31} - \mu\rho_{21})]} \right|^2.
\end{eqnarray}
\end{widetext}
Here, it is worthwhile to note that the denominator of Eq.~(\ref{rho31+mu rho32}) is nothing else than the determinant $D$ of the system of Eqs.~(\ref{R31 extended}) and~(\ref{R32 extended}) in the steady-state regime. The zeros of $D$ determine eigen frequencies (in the rotating frame) $\kappa_\pm$ of two dressed $\Lambda$-emitter states (see Sec.~\ref{Discussion}). Being a quadratic in $\Delta_{31}$ function (recall that $\Delta_{32} = \Delta_{31} - \Delta_{32}$), $D$ can be written as $D = (\Delta_{31} - \kappa_{+})(\Delta_{31} - \kappa_{-})$. Importantly, $\kappa_{\pm}$ are functions of the doublet splitting $\Delta_{21}$ and the Rabi magnitude $|\Omega_0|$ of the incident field and are obtained after solving the whole nonlinear steady-state problem.

Figure~\ref{fig:kappa plus} shows trajectories ($|\Omega_0|^2$-dependence) of $\kappa_\pm$ (solid curves) calculated for two values of the detuning $\Delta_{31} = 1125$ (left plot) and $\Delta_{31} = 1200$ (right plot), setting the doublet splitting $\Delta_{21} = 15$ and relaxation rate $\gamma_{21} = 0.01$. The corresponding $R$-vs-$|\Omega_0|^2$ dependencies (scaled to plots) are shown as dashed curves. Observing left plot, we come to the conclusion that a peak in reflection occurs when $\mathrm{Re}[\kappa_+]$ crosses horizontal line ($\mathrm{ Re}[\kappa_+] = \Delta_{31}$), i.e. when the system is in resonance with the incident field. At $\Delta_{31} = 1125$, the crossing takes place twice, thus bringing two peaks to the reflectance $R$. The fact that the peak value of $R$ is nearly unity (total reflection) can be qualitatively explained as follows. First, keep in mind that the incident and secondary fields, being in resonance with each other, have exactly opposite phases, in such a way interfering destructively. Next, both field will interfere out completely, giving the null transmitted field [$\Omega_\mathrm{tr} = \Omega_0 + \gamma_R (\rho_{31} + \mu\rho_{32}) = 0$] when their amplitudes are equal to each other ($|\Omega_0| = \gamma_R |\rho_{31} + \mu\rho_{32}|$). This condition for $\gamma_R = 100$ is fulfilled for both $R$ peaks.

 In the case of $\Delta_{31} = 1200$ (Fig.~\ref{fig:kappa plus}, right plot) the trajectory of $\mathrm{Re}[\kappa_+]$ does not cross $\Delta_{31}$-line, resulting in only one peak in reflection. As $\mathrm{Re}[\kappa_+] \neq \Delta_{31}$, the incident and secondary fields are not exactly in unti-phase and thus can not completely compensate each other. Consequently, the peak value of $R$ is visibly smaller than unity (see Fig.~\ref{fig:Reflectance Delta21 = 100}, upper-left plot).

We do not show the results for $R$ in the vicinity of the second root $\kappa_{-}$ because reflectance is low there.

\subsection{Time-domain}
\label{Reflectance dynamics}
As was found in the preceding section, within some range of the incident field intensity $|\Omega_0|^2$, which depends on the detuning $\Delta_{31}$, the reflectance $R$ is unstable (see Fig.~\ref{fig:Reflectance Delta21 = 100}). Surprisingly, the instability may develop at relatively low values of $|\Omega_0|^2 \lesssim 1$, as it takes place, in particular, for $\Delta_{31} = 1000$.

 To uncover the character of instabilities, we performed time-domain calculations of the reflectance $R$ within the unstable region, basically interesting in the system behavior at $|\Omega_0|^2 \sim 1$. The typical results are shown in Fig.~\ref{fig:Reflectance dynamics Delta21 = 15}. The set of parameters used in calculations are specified in the figure caption.

As observed from Fig.~\ref{fig:Reflectance dynamics Delta21 = 15}, the only type of attractors, exhibiting by the system in the range of $|\Omega_0|^2 \sim 1$, is self-oscillations, which is derived from the equidistant character of the Fourier spectrum $|\int_{T} \exp{i\omega t} \Omega_\mathrm{refl}(t) dt|$ (middle panels) and
closedness of the two-dimensional phase-space map ($\mathrm{Re}[\Omega_\mathrm{refl}], \mathrm{Im}[\Omega_\mathrm{refl}]$) of the attractor

It is worth to noting that the Fourier spectrum of self-oscillations contains components residing in THz domain (see panels b) and d) in Fig.~\ref{fig:Reflectance dynamics Delta21 = 15}). Thus, in this regime, the system represents a source of coherent THz radiation.

\section{Summary}
\label{Summary}
We have conducted a theoretical study of the optical response of a quantum metasurface comprizing regularly spaced $\Lambda$-emitters subjected to a CW quasi-resonant excitation. The (secondary) field, acting on an emitter on the part of the others, has been taken into account within the framework of the mean-field approximation. This field, being dependent on the emitter current state, plays the role of a positive feedback. An exact method of solving the nonlinear steady-state problem has been put forward to unravel the multi-valued character of the system's optical response. Using the Lyapunov's exponent analysis, we have found windows of stability and instability of different branches of the monolayer optical response.

Furthermore, the bifurcation diagram of the response has been put forward in order to get a general insight into possible scenarios of the system motion. Different types of bifurcations, such as supercritical and subcritical Andronov-Hopf, limit cycle--chaos bifurcations, has been found.

The monolayer optical response has turned out to manifest very different dynamics under a CW excitation: periodic self-oscillations and chaotic behavior. Herewith, the frequency of self-oscillations has turned out to depend on the incident field magnitude and, for the set of parameters used, has fall into the THz domain.
The instabilities found originate from the complicated interplay of the secondary field--driven dynamic dressing of the emitter's transitions and population redistribution among the dressed states.

Within the frequency range around the collective (excitonic) resonance, the monolayer has been found to almost totally reflect the incident field, thus acting as a perfect mirror. On the top of it, the monolayer's reflectance can be abruptly switched from a low to high level and back by a small variation of the driving field magnitude, implying bistability and hysteresis. Note that an atomically thin layer of MoSe$_2$ manifests similar behavior~\cite{Back2018,Scuri2018}. Likewise, quantum metasurfaces of arrays of atoms trapped in optical lattices~\cite{BekensteinNatPhys2020} also exhibit high reflectance. So our system is a representative of one more class of nanoscale objects of such a type. Importantly, however, the advantage of metasurfaces comprizing, in particular, SQDs as quantum emitters, infers from possibility of controlling properties of the SQD-based metasurfaces by the geometry and materials of the nanostructure.

 The results obtained suggest various practical applications of metasurfases of quauntum $\Lambda$-emitters as being a nanometer-thin bistable mirror, a tunable generator of coherent THz radiation (in self-oscillation regime), and an optical noise generator (in chaotic regime). The intrinsic sensitivity of the optical response to the initial conditions in the chaotic regime could be of interest for information encryption~\cite{Gao2008}. All these findings make the considered system a promising candidate for all-optical information processing and computing.
\\
\\
{\bf Notes}
\\
\noindent
The authors declare no competing financial interest.
\\

\acknowledgments
R. F. M. acknowledges M. Akmullah Bashkir State Pedagogical University for a financial support.
A. V. M. acknowledges support from Spanish MINECO grant MAT2016-75955.
%



\begin{thebibliography}{99}
%
\bibitem{Novoselov2004} K. S. Novoselov, A. K. Geim, S. V. Morozov, D. Jiang, Y. Zhang, S. V. Dubonos, I. V. Grigorieva, A. A. Firsov,
    Electric field effect in atomically thin carbon films,
    Science \textbf{306}, 666 (2004).
%
\bibitem{CastroNeto2009} A. H. Castro Neto, F. Guinea, N. M. R. Peres, K. S. Novoselov, and A. K. Geim,
    The electronic properties of graphene,
    Rev. Mod. Phys. \textbf{81}, 109 (2009).
%
\bibitem{ManzeliNatMatRev2017} S. Manzeli, D. Ovchinnikov, D. Pasquier, O. V. Yazyev and A. Kis,
    2D transition metal dichalcogenides,
    Nat. Mat. Rev. \textbf{2}, 17033 (2017).
%
\bibitem{ChernozatonskiiPhysicsUsp2018} L. A. Chernozatonskii, A. A. Artyukh,
    Quasi-two-dimensional transition metal dichalcogenides:
    structure, synthesis, properties, and applications,
    Physics-Usp \textbf{61}, 2 (2018).
%
\bibitem{BonaccorsoMaterialsToday2012} F. Bonaccorso, A. Lombardo, T. Hasan, Z. Sun, L. Colombo, and A. C. Ferrari,
    Production and processing of graphene and 2d,
    Materials Today \textbf{15}, 564 (2012).
%
\bibitem{BhimanapatiACSNano2015}
G. R. Bhimanapati, Z. Lin, V. Meunier et al.,
    Recent advances in two-dimensional materials beyond graphene,
    ACS Nano \textbf{9}, 11509 (2015).
%
\bibitem{TanChemRev2017} C. Tan, X. Cao, X.-J. Wu, Q. He, J. Yang, X. Zhang, J. Chen, W. Zhao, S. Han, G.-H. Nam, M. Sindoro, and H. Zhang,
    Recent advances in ultrathin two-dimensional nanomaterials,
    Chem. Rev. \textbf{117}, 6225 (2017).
%
\bibitem{JariwalaNatMat2017} D. Jariwala, T. J. Marks and M. C. Hersam,
    Mixed-dimensional Van Der Waals Heterostructures,
    Nat. Mat. \textbf{16}, 170 (2017).
%
\bibitem{MuNanoStrucNanoObjects2018} P. Mu, G. Zhou, C.-L. Chen,
    2D nanomaterials assembled from sequence-defined molecules,
    Nano-Struc. Nano-Objects \textbf{15}, 153 (2018).
%
\bibitem{Evers2013} W. H. Evers, B. Goris, S. Bals, M. Casavola, J. de Graaf, R. van Roij, M. Dijkstra, and D. Vanmaekelbergh,
    Low-dimensional semiconductor superlattices formed by geometric control over nanocrystal attachment,
    Nano Lett. \textbf{13}, 2317 (2013).
%
\bibitem{Baranov2015} A. V. Baranov, E. V. Ushakova, V. V. Golubkov, A. P. Litvin
P. S. Parfenov, A. V. Fedorov, and K. Berwick,
    Self-organization of colloidal PbS quantum dots into highly ordered superlattices,
    Langmuir \textbf{31}, 506 (2015).
%
\bibitem{Ushakova2016} E. V. Ushakova, S. A. Cherevkov, A. P. Litvin, P. S. Parfenov,
D.-O. A. Volgina, I. A. Kasatkin, A. V. Fedorov, and A. V. Baranov,
    Ligand-dependent morphology and optical properties of lead sulfide quantum dot superlattices,
    J. Phys. Chem. C  \textbf{120}, 25061 (2016).
%
\bibitem{Liu2017} W. Liu, X. Luo, Y. Bao, Y. P. Liu, G.-H. Ning, I. Abdelwahab, L. Li, C. T. Nai, Z. G. Hu, D. Zhao, B. Liu, S. Y. Quek, K. P. Loh,
        A two-dimensional conjugated aromatic polymer via C-C coupling reaction,
        Nat. Chem. \textbf{9}, 563 (2017)..
%
\bibitem{BaimuratovSciRep2013} A. S. Baimuratov, I. D. Rukhlenko, V. K. Turkov, A. V. Baranov, A. V. Fedorov,
    Quantum-dot supercrystals for future nanophotonics,
    Sci. Rep. \textbf{3}, 1727 (2013).
%
\bibitem{BaimuratovOptLett2013} A. S. Baimuratov, I. D. Rukhlenko, and A. V. Fedorov,
    Engineering band structure in nanoscale quantum-dot supercrystals,
    Opt. Lett. \textbf{38}, 2259 (2013).
%
\bibitem{BaimuratovOptLett2017} A. S. Baimuratov, A. I. Shlykov, W. Zhu, M. Yu. Leonov, A. V. Baranov, A. V. Fedorov, and I. D. Rukhlenko,
    Excitons in gyrotropic quantum-dot supercrystals,
    Opt. Lett. \textbf{42}, 2423 (2017).
%
\bibitem{VovkPhysChemChemPhys2018} I .A. Vovk,  N. V. Tepliakov, A. S. Baimuratov, M. Yu. Leonov,  A. V. Baranov, A. V. Fedorov and  I. D. Rukhlenko,
    Excitonic phenomena in perovskite quantum-dot supercrystals,
    Phys. Chem. Chem. Phys. \textbf{20}, 25023 (2018).
%
\bibitem{NosseMicroelectronJourn2008} J. F. Nossa and A. S. Camacho,
    Optical properties of supercrystals,
    Microelectronics Journal \textbf{39}, 1251 (2008)
%
%
\bibitem{BaimuratovSciRep2016} A. S. Baimuratov, Y. K. Gun’ko, A. V. Baranov, A. V. Fedorov, I. D. Rukhlenko,
    Chiral quantum supercrystals with total dissymmetry of optical response,
    Sci. Rep. \textbf{6}, 23321 (2016).
%
\bibitem{MalikovEPJWebConf2017} R. F. Malikov, I. V. Ryzhov, and V. A. Malyshev,
    Nonlinear optical response of a 2D quantum dot supercrystal,
    EPJ Web of conferences \textbf{161}, 02014 (2017).
%
\bibitem{MalyshevJPhysConfSer2019} V. A. Malyshev, P. \'A. Zapatero, A. V. Malyshev, R. F. Malikov, and I. V. Ryzhov,
    Nonlinear optical dynamics of a 2D semiconductor quantum dot super-crystal: Emerging multistability, self-oscillations and chaos,
    J. Phys. Conf. Ser. \textbf{1220}, 012006 (2019).
%
\bibitem{RyzhovPRA2019} I. V. Ryzhov, R. F. Malikov, A. V. Malyshev, and V. A. Malyshev,
    Nonlinear optical response of a two-dimensional quantum-dot supercrystal: Emerging multistability, periodic and aperiodic self-oscillations, chaos, and transient chaos,
    Phys. Rev. A \textbf{100}, 003800 (2019).
%
\bibitem{VlasovJApplSpectrosc2013} R. A. Vlasov, A. M. Lemeza, M. G. Gladush,
    Resonance fluorescence of optically dense ensembles of three-level resonant centers under conditions of energy-level population auto-oscillations,
    J. Appl. Spectrosc. {\bf 80}, 698 (2013).
%
\bibitem{VlasovLasPhysLett2013} R. A. Vlasov, A. M. Lemeza, M. G. Gladush,
    Dynamical instabilities of spectroscopic transitions in dense resonant media,
    Las. Phys. Lett. \textbf{10}, 045401 (2013).
%
\bibitem{BayramdurdiyevEPJWebConf2019}
D. Bayramdurdiyev, R. Malikov,  I. Ryzhov, V. Malyshev,
    Multistability and high reflectance of a mono-layer of three-level quantum emitters with a doublet in the excited state,
    EPJ Web of Conferences \textbf{220}, 03004 (2019).
%
\bibitem{BayramdurdiyevJETP2020}
D. Y. Bayramdurdiyev, R. F. Malikov,  I. V. Ryzhov, V. A. Malyshev,
    Nonlinear optical dynamics and high reflectance of a monolayer of three-level quantum emitters with a doublet in the excited state,
    J. Exp. Theor. Phys. \textbf{131}, 244 (2020).
%
\bibitem{Brunner2009} D. Brunner, B. D. Gerardot, P. A. Dalgarno, G. Wüst, K. Karrai, N. G. Stoltz, P. M. Petroff, R. J. Warburton,
    A coherent single-hole spin in a semiconductor,
    Science \textbf{325}, 70 (2009).
%
\bibitem{BookNanocrystal2011} K. Baba, H. Kasai, K. Nishida, and H. Nakanishi,
    Functional organic nanocrystals,
    in {\it Nanocrystals}, ed. Y. Masuda (IntechOpen, 2011) Ch 15, p. 397.
%
\bibitem{Benedict1990} M. G. Benedict, A. I. Zaitsev, V. A. Malyshev, and E. D. Trifonov,
    Mirrorless bistability in transmission of ultra short light pulses through a thin layer with resonant two-level centers,
    Opt. Spectrosc. \textbf{68}, 473 (1990).
%
\bibitem{Benedict1991} M. G. Benedict, A. I. Zaitsev, V. A. Malyshev, and E. D. Trifonov,
    Reflection and transmission of ultrashort light pulses through a thin resonant medium: Local-field effects,
    Phys. Rev. A \textbf{43}, 3845 (1991).
%
\bibitem{Bowden1986} Y. Ben-Aryeh, C. M. Bowden, and J. C. Englund,
    Intrinsic optical bistability in collections of spatially distributed two-mevel atoms,
    Phys. Rev. A \textbf{34}, 3917 (1086).
%
%
\bibitem{Basharov1988} A. M. Basharov,
    Thin film of two-level atoms: a simple model of optical bistability and self-pulsation,
    Sov. Phys. JETP \textbf{67}, 1741 (1988).
%
\bibitem{Oraevsky1994} A. N. Oraevsky, D. J. Jones, and D. K. Bandy,
    Semiconductor microballs as bistable optical elements,
    Opt. Commun. \textbf{111}, 163 (1994).
%
\bibitem{Malyshev2000} V. A. Malyshev and E. Conejero Jarque,
    A thin film of oriented short linear Frenkel chains as an optical bistable element,
    Opt. Express \textbf{6}, 227 (2000).
%
\bibitem{Glaeske2000} H. Glaeske, V. A. Malyshev, and K.-H. Feller,
    Intrinsic optical bistability of ultrathin film consisiting of oriented linear aggregates,
    J. Chem. Phys. \textbf{113}, 1170 (2000).
%
\bibitem{Klugkist2007} J. A. Klugkist, V. A. Malyshev, and J. Knoester,
    Intrinsic optical bistability of thin films of linear
    molecular aggregates: The one-exciton approximation,
    J. Chem. Phys. \textbf{127}, 164705 (2007)
%
\bibitem{Malikov2017} R. F. Malikov and V. A. Malyshev,
    Optical bistability and hysteresis of a thin slab of resonant emitters: Interplay of inhomogeneous broadening of the absorption line and the Lorentz local field,
    Opt. Spectrosc. \textbf{122}, 955 (2017).
%
\bibitem{EckmannRevModPhys1985} J.-P. Eckmann and D. Ruelle,
    Ergodic theory of chaos and strange attractors,
    Rev. Mod. Phys. \textbf{57}, 617 (1985).
%
\bibitem{AndronovBook1966} A. A. Andronov, A. A. Vitt, S. E. Khaikin, {\it Theory Of Oscillators} (Pergamon Press, New York, 1966).
%
\bibitem{GuckenheimerBook1986} J. Guckenheimer, P. Holmes, {\it Nonlinear Oscillations, Dynamical Systems and Bifurcations of Vector Fields}, Second
Printing (Springer, Berlin, 1986).
%
\bibitem{NeimarkLandaBook1992} Yu. I. Neimark and P. S. Landa, {\it Stochastic and Chaotic Oscillations} (Springer Science\&Bussiness Media, 1992).
%
\bibitem{OttBook1993} E. Ott, {\it Chaos in Dynamical Systems} (Cambridge University Press, Cambridge, 1993).
%
\bibitem{Arnol'dBook1994} V. I. Arnol'd (Ed.), V. S. Afrajmovich, Yu. S. Il'yashenko, L. P. Shil'nikov,
    {\it Dynamical Systems V: Bifurcation Theory and Catastrophe Theory} (Springer, 1994).
%
\bibitem{AlligoodBook1996} K. T. Alligood, T. D. Sauer, and J. A. Yorke, {\it Chaos: An Introduction to Dynamical Systems} (Springer, 1996).
%
\bibitem{KatokBook1997} A. Katok and B. Hasselblatt, {\it Introduction to the Modern
    Theory of Dynamical Systems} (Cambridge University Press, 1997).
%
\bibitem{KuznetsovBook2004} Yu. A. Kuznetsov, {\it Elements of Applied Bifurcation Theory}, 3rd edition (Springer, 2004).
%
\bibitem{RyzhovEPJWebConf2019}
I. Ryzhov, R. Malikov, A. Malyshev, V. Malyshev,
    A monolayer of three-level quantum $\Lambda$-emitters: A perspective system from the viewpoint of nonlinear optical dynamics and nanophotonics,
    EPJ Web of Conferences \textbf{220}, 02012 (2019).
%
\bibitem{Lindblad1976} G. Lindblad,
    On the generators of quantum dynamical semigroups,
    Commun. Math. Phys. \textbf{48}, 119 (1976).
%
\bibitem{BlumBook2012} K. Blum, {\it Density Matrix: Theory and applications}, 3rd edition (Springer, 2012).
%
\bibitem{BornAndWolf} M. Born and E. Wolf, {\it Principles of Optics}, 6-th edition (Springer, 1980).
%
\bibitem{Dicke1954} R. H. Dicke.
    Coherence in Spontaneous Radiation Processes,
    Phys. Rev. \textbf{93}, 99 (1954).
%
\bibitem{MalikovJETP1979} E. D. Trifonov, A. I. Zaitsev, R. F. Malikov,
    Superradiance of an extended system,
    Sov. Phys. JETP \textbf{49}, 33 (1979).
%
\bibitem{BenedictBook1996} M. G. Benedict, A. M. Ermolaev, V. A. Malyshev, I. V. Sokolov, E. D. Trifonov,
    {\it Super-radiance: Multiatomic Coherent Emission} (IOP Publishing, Bristol, 1996).
%
\bibitem{Lai2011} Y.-C. Lai and T. T\'el, {\it Transient Chaos. Complex dynamics in finite-time scales} (Springer, Berlin, 2011).
%
\bibitem{Tel2015} T. T\'el,
    The joy of transient chaos,
    Chaos \textbf{25}, 097619 (2015),
%
\bibitem{Back2018} P. Back, S. Zeytinoglu, A. Ijaz, M. Kroner, and A. Imamo\u{g}lu,
    Realization of an electrically tunable narrow-bandwidth atomically thin mirror using monolayer MoSe,
    Phys. Rev. Lett. \textbf{120}, 037401 (2018).
%
\bibitem{Scuri2018} G. Scuri, Y. Zhou, A. A. High, D. S.Wild, C. Shu, K. De Greve, L. A. Jauregui, T. Taniguchi, K. Watanabe, P. Kim, M. D. Lukin, and H. Park,
    Large excitonic reflectivity of monolayer MoSe2 encapsulated in hexagonal boron nitride,
    Phys. Rev. Lett. \textbf{120}, 037402 (2018).
\bibitem{BekensteinNatPhys2020} R. Bekenstein, I. Pikovski, H. Pichler, E. Shahmoon, S. F. Yelin, and M. D. Lukin,
    Quantum metasurfaces with atom arrays,
    Nat. Physics \textbf{16}, 676 (2020).
%
\bibitem{Gao2008} T. Gao, Z. Chen,
A new image encryption algorithm based on hyper-chaos,
    Phys. Lett. A \textbf{372}, 394 (2008).
%
\end{thebibliography}


\end{document}